\documentclass[%
reprint,
 showpacs,preprintnumbers,
 bibnotes,
 amsmath,amssymb,
 aps,
 prb,
floatfix
]{revtex4-1}

\usepackage[utf8]{inputenc}

\usepackage{graphicx}
\usepackage{dcolumn}
\usepackage{bm}
\usepackage{epstopdf}
\usepackage{color}

\usepackage{color}
\usepackage{float}
\usepackage{amsfonts}
\usepackage{mathtools}
\usepackage{amsmath}
\usepackage{amssymb}

\begin{document}


\title{Spin-valley system in a gated MoS$_2$-monolayer quantum dot}
\author{J. Paw\l{}owski}
\email[]{jaroslaw.pawlowski@pwr.edu.pl}
\affiliation{
Department of Theoretical  Physics, Wroc\l{}aw University of Science and Technology, Wybrze\.{z}e Wyspia\'{n}skiego 27, 50-370 Wroc\l{}aw, Poland}

\date{\today}

\begin{abstract}
The aim of presented research is to design a nanodevice based on a gate-defined quantum dot within a MoS$_2$  monolayer in which we confine a single electron. By applying control voltages to the device gates we modulate the confinement potential and force intervalley transitions. The present Rashba spin-orbit coupling additionally allows for spin operations. Moreover, both effects enable the spin-valley SWAP. The device structure is modeled realistically, taking into account feasible dot-forming potential and electric field that controls the Rasha coupling. Therefore, by performing reliable numerical simulations, we show how by electrically controlling the state of the electron in the device, we can obtain single- and two-qubit gates in a spin-valley two-qubit system. Through simulations we investigate possibility of implementation of two qubits \textit{locally}, based on single electron, with an intriguing feature that two-qubit gates are easier to realize than single ones.
\end{abstract}

\pacs{73.63.Kv, 73.22.-f, 85.35.Gv, 03.67.Lx}

\maketitle


\section{Introduction}
Two-dimensional crystals consisting of single layers of atoms are modern materials that can be used for implementation of quantum computation. 2D monolayers of transition metal dichalcogenides (TMDCs), e.g. MoS$_2$, seem to be better candidates than graphene because of their wide band gaps and strong electrically induced spin-orbit coupling of the Rashba type\cite{strano,prx}. By considering the valley degree of freedom of an electron together with its spin we extend our ability to define a qubit into: spin, valley\cite{mymos2} and hybrid spin-valley qubit\cite{palyi}. However, the most interesting is the definition based on spin and valley of a single electron as a two-qubit system\cite{rohling,wu}. 

The area of application of monolayer materials for construction of electronic nanodevices is currently under strong development
\cite{nano4,nano5,nano2,strano,nano7, nano11,nano12}.
Methods for building devices based on gated TMDC monolayers or nanotubes become increasingly advanced\cite{nano1,nano3,wang,nano6,nano10}, opening the possibility of utilizing the spin and valley index of electrons controlled therein.
In particular, it is shown by recent results with electrostatic quantum dots (QDs) with a gated MoS$_2$-nanoribbon-QD measured by a single electron transport\cite{nano,zhuo}, or tunable TMDC spintronic devices, where spin or valley polarized currents emerge in TMDC monolayer proximitized by nearby ferromagnetic\cite{oscar1,fabiann,yuan,shao,leyla}. Transistor structures with TMDC monolayer forming active area in tunnel FETs are being developed\cite{mono}, also with vertical TMDCs heterostructures \cite{vhetero0,vhetero}. The more intriguing lateral, in-plane TMDCs heterojunctions are also constructed, enabling interesting 1D physics at interfaces\cite{oscar2}, or leading to improved FETs switching characteristics \cite{hetero1,hetero2,hetero3}.

Inspired by this, we examined the possibility of realization of a nanodevice based on a MoS$_2$ monolayer, capable of creating a two-qubit system defined on spin and valley degrees of freedom of a confined electron. For this purpose, we have built a realistic model of the nanodevice and perform numerical simulations that prove its capabilities. 
Thanks to the use of appropriately modulated local control voltages, the system is all-electrically controlled and does not require using photons or external microwaves, 
thus significantly improving its scalability.

\section{Model}

In this section we will go through the device model. The potential in the entire nanodevice, controlled by the gate voltages, is calculated by solving the Poisson equation, while the electron states in the flake are described with the tight-binding formalism. Let's start with the nanostructure overview.

\subsection{Device structure}

\begin{figure}[b]
	\center
	\includegraphics[width=8cm]{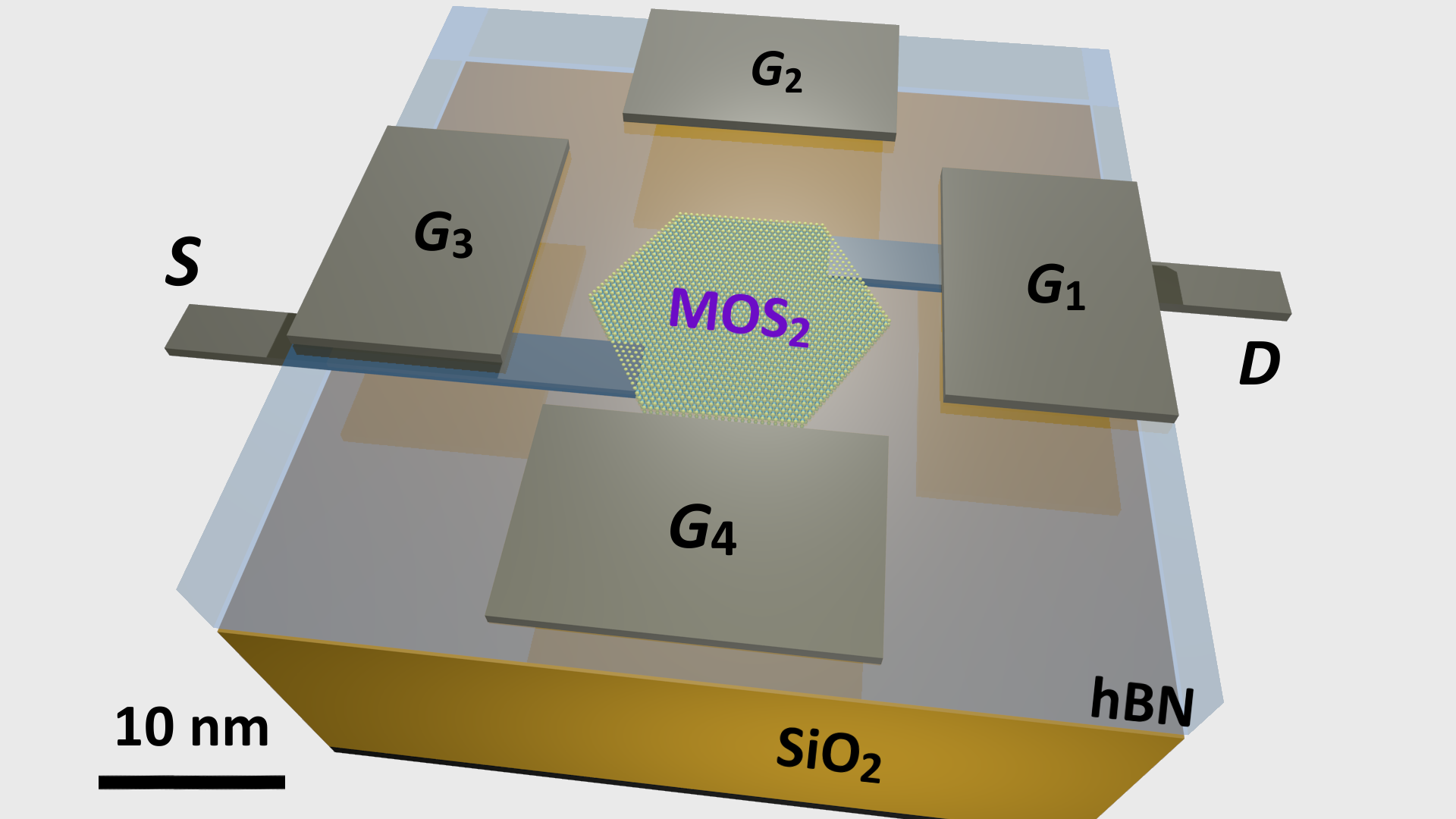}
	\caption{\label{fig:1} The proposed nanodevice structure containing a MoS$_2$ monolayered flake deposited on a SiO$_2$ layer, and separated by hBN from the control gates responsible for creating the QD confinement potential.}
\end{figure}

The proposed nanodevice structure is presented in Fig.~\ref{fig:1}. On a strongly doped silicon substrate we place a 20-nm-thick layer of SiO$_2$. Then we place two electrodes which serve as a source ($S$) and a drain ($D$). Directly on them we deposit a MoS$_2$-monolayer (hexagonally shaped) flake of 16-nm-diameter. The monolayer is then covered with a $5$-nm thick insulating layer of hexagonal boron nitride (hBN) with a large bandgap\cite{hbn}, forming a tunnel barrier. Finally on top of the sandwiched structure we lay down four 15-nm-wide control gates ($G_{1..4}$), placed symmetrically around the central square-like gap of size $20\times20$~nm. 
The gate layout presented here is quite similar to the one proposed by us recently\cite{mymos2}, but with a larger $20$~nm clearance between opposite gates, which may ease their deposition.

\begin{figure}[h]
	\center
	\includegraphics[width=8cm]{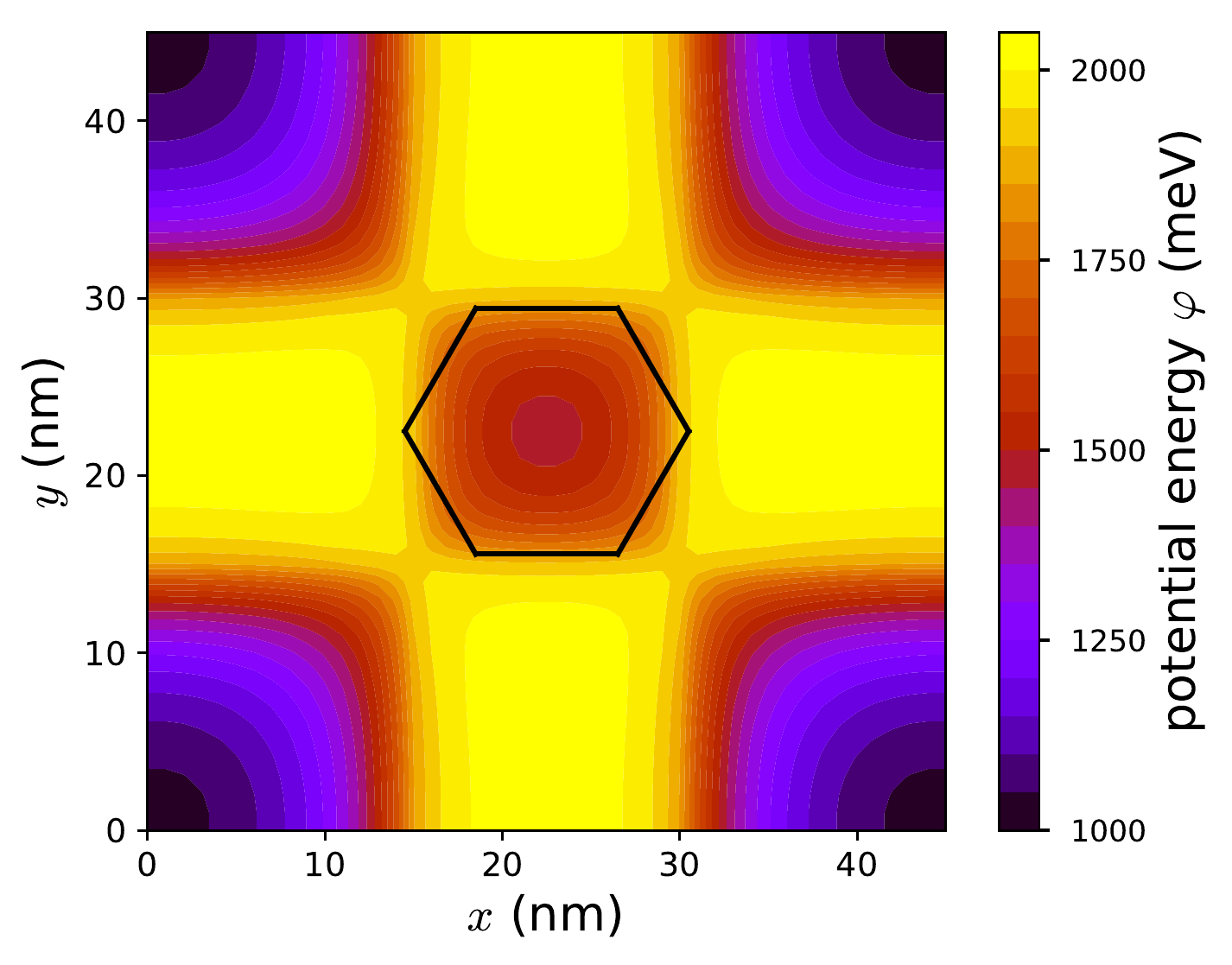}
	\caption{\label{fig:2} The confinement potential energy at the area where the monolayer lies (indicated by a black hexagon), created by the four symmetrically arranged control gates (see Fig.~\ref{fig:1}).}
\end{figure}

Source, drain and the gates layout are clearly presented in Fig.~\ref{fig:1}.
Voltages applied to these gates (relative to the substrate) are used to create confinement in the flake. To calculate realistic electrostatic potential $\phi(\mathbf{r})$ we solve the Poisson equation taking into account voltages $V_{1..4}$ applied to control gates $G_{1..4}$ and to the highly doped substrate $V_0=0$, together with space-dependent permittivity of different materials in the device\cite{mymos2,mydrut}. Resulting potential in the area between SiO$_2$ and hBN layers, where the flake is sandwiched, is presented in Fig.~\ref{fig:2}.
We deplete the electron gas until a single electron remains in the formed dot confinement potential.

\subsection{Monolayer model}

The monolayer flake is made of molybdenum disulfide.
\begin{figure}[h]
	\center
	\includegraphics[width=8cm]{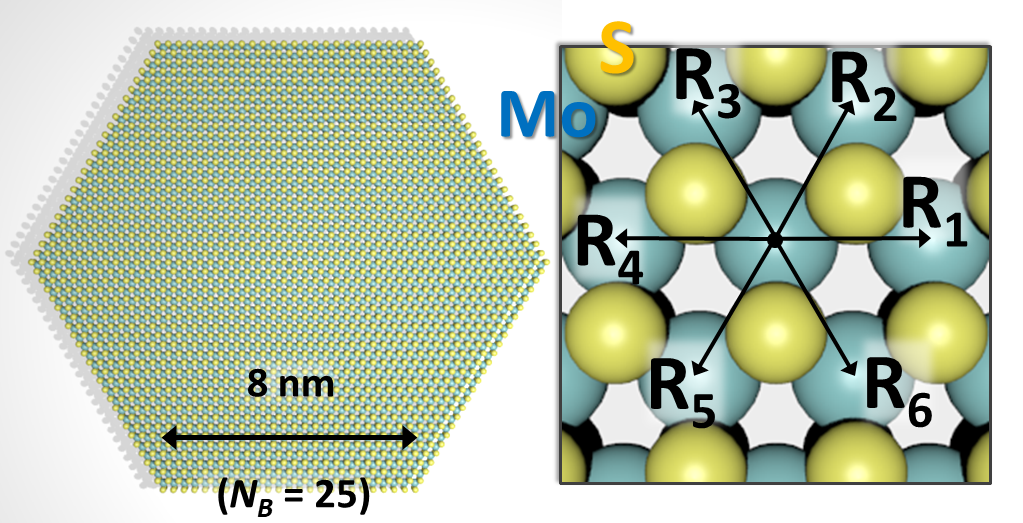}
	\caption{\label{fig:3} The MoS$_2$ monolayered flake structure: (left) hexagonal flake employed in the device has sides made of $N_B=25$ Mo atoms, giving the flake side of $8$~nm (lattice distance between Mo nodes is $0.319$~nm); (right) MoS$_2$ crystal lattice structure formed of hexagonally packed Mo and S atoms arranged in triangular lattices rotated relative to each other by $\pi$. The Mo lattice vectors $R_k$ determine the hopping directions in our nearest-neighbors TB model.}
\end{figure}
MoS$_2$ monolayers are successfully described by several tight-binding (TB) models, with different numbers of orbitals used, including nearest or next-nearest neighbors. Seven\cite{azgari} or eleven\cite{ridolfi,capp,fang} Mo and S orbitals construct TB basis to reproduce low-energy physics in the entire Brillouin zone, also near the $\mathrm{\Gamma}$-point. Although the simpler three-band (including three Mo orbitals) TB model\cite{xiao} fails around the $\mathrm{\Gamma}$-point, it correctly represents the orbital composition around the K point near the (both conduction and valence) band edges, where the Bloch states mainly consist of Mo $d$ orbitals\cite{haw}. 
Thus it is good enough to deal with low energy states near the band minimum.
However when considering perpendicular electric field, crucial for the Rashba coupling, we have to include also S orbitals localized above and below Mo plane in the MoS$_2$ structure (see Fig.~\ref{fig:3}). For calculating Rashba coupling we will utilize 11-band model with three $p$ orbitals for each S atom in dimer.\cite{ridolfi}

Consequently, we have described the monolayer structure using three Mo orbitals:  $d_{z^2}$, $d_{xy}$, $d_{x^2-y^2}$, and the nearest-neighbors hoppings\cite{xiao}:
\begin{align}\label{ham1}
H=&\sum_{i\,\sigma \sigma'\alpha \beta }\delta_{\sigma \sigma'}\delta_{\alpha \beta}(\epsilon_{\alpha }+{\varphi }_i)\,{\hat{n}}_{i\alpha \sigma }+{s}^z_{\sigma \sigma'}{\lambda }_{\alpha \beta }\,{\hat{c}}^{\dagger }_{i\alpha \sigma }{\hat{c}}_{i\beta \sigma '}\nonumber\\
+&\sum_{\langle ij\rangle\,\sigma \sigma'\alpha \beta }{\delta_{\sigma \sigma'}t_{\alpha \beta }\,{\hat{c}}^{\dagger }_{i\alpha \sigma }{\hat{c}}_{j\beta \sigma }}
+H_R+H_Z.
\end{align}
The potential energy of the electrostatic confinement at the $i$-th lattice site: $\varphi_i=-|e|\phi(x_i,y_i)$ together with the on-site energies $\epsilon_\alpha$ enter the  on-diagonal matrix elements ($\alpha$ numbers the orbitals).

The off-diagonal electron hopping element from the $\beta$ Mo orbital localized in the $j$-th lattice site to the $\alpha$ orbital localized in the $i$-th site is denoted by $t_{\alpha \beta }\equiv t_{\alpha \beta }\!\left(R_{k\left(i,j\right)}\right)$. It depends on the hopping direction (between $\langle ij\rangle$ neighbor pair) described by the nearest neighbor vectors $R_k$ for the molybdenum (Mo) lattice, which are defined as in Fig.~\ref{fig:3}. They form two non-equivalent families: $R_1$, $R_3$, $R_5$ and $R_2$, $R_4$, $R_6$ with the nearest sulphur (S) neighbor on the left or right side, as shown in Fig.~\ref{fig:3}. This symmetry constraint reflects on the reciprocal lattice where in the corners of the first (hexagonal) Brillouin zone, the K points form two non-equivalent families: $K$ and $K’$.

Opposite hoppings are mutually transposed: $t_{\alpha \beta }(R_k)=t_{\beta \alpha }(-R_k)$. Their explicit forms, together with the on-site energies $\epsilon_\alpha$, can be found in [\onlinecite{mymos2,xiao}].

\subsection{Rashba coupling}

Electric field perpendicular to the monolayer surface breaks the reflection $\sigma_h$ symmetry and modifies the on-site energies of atoms in three MoS$_2$ sublayers. This leads to externally, electrically controlled spin-orbit interaction (SOI) of the Rashba type. The Rashba coupling can be also introduced to layered TMDCs by a structure asymmetry from ferromagnetic substrate leading to the proximity effect \cite{oscar1}.

The idea to calculate the electrically induced Rashba spin-orbit coupling strength is to take the tight-binding model with atomic spin-orbit coupling (introduced by $\lambda_{\alpha\beta}$ in Eq.~\ref{ham1}) including also $p$ orbitals of the sulfur top and bottom sublayers to which we apply on-site potentials $V_\mathrm{t}$ and $V_\mathrm{b}$\cite{down1,azgari,down3,down4}.
The difference between them results from external electric field:  $V_\mathrm{b}(x,y)=E_z(x,y,0)d/2$, $V_\mathrm{t}(x,y)=-V_\mathrm{b}(x,y)$.
While  $d=0.32$~nm is the monolayer thickness (sulfur sublayers distance). 

Given such extended tigh-binding model (we take 11-orbital model of Ridolfi et al. [\onlinecite{ridolfi}]), we perform downfolding, using the L\"owdin partitioning technique, to our Mo-orbitals model and obtain the Rashba coupling $\gamma_R$ within this 3-band base. Further details of the calculation are attached in the Appendix. Resulting coupling matrix elements are proportional to external electric field: $\gamma^{\alpha\beta}_R(x,y)=|e|E_z(x,y)\,\gamma_{\alpha\beta}$, with the explicit form
\begin{equation}
\label{gamm}
\gamma=
\begin{pmatrix}
0.1 & 5 & 17\\
5 & 4 & 1\\
17 & 1 & 4
\end{pmatrix}
\times 10^{-3}\;\mathrm{nm}.
\end{equation}

The tight-binding Rashba Hamiltonian\cite{rb1,rb2,down4,rb3,rb4}  (in Eq.~\ref{ham1}) with characteristic spin- and orientation-dependent hopping between two nearest neighbor bonds is:  
\begin{equation}
\label{rash}
H_R=\sum_{\langle ij\rangle\,\sigma \sigma'\alpha \beta } \imath\gamma^{\alpha\beta}_R  (\hat{\mathbf{e}}_{ij}\! \times \hat{\mathbf{z}})\!\cdot \mathbf{s}_{\sigma\sigma'}\, c^\dagger_{i\alpha\sigma}c_{j\beta\sigma'},
\end{equation}
with the Pauli-matrices vector $\mathbf{s}=(\sigma_x,\sigma_y,\sigma_z)$, and the (unit) versor $\hat{\mathbf{e}}_{ij}\equiv(e^x_{ij},e^y_{ij})$ pointing along the bond connecting sites $i$ and $j$. An obvious property $\hat{\mathbf{e}}_{ji}=-\hat{\mathbf{e}}_{ij}$ ensures $H_R$ hermiticity. The expanded hopping expression is $(\hat{\mathbf{e}}_{ij}\! \times \hat{\mathbf{z}})\!\cdot \mathbf{s}_{\sigma\sigma'}=e^y_{ij} s^x_{\sigma\sigma'} - e^x_{ij} s^y_{\sigma\sigma'}$, where $s^x=\sigma_x$, $s^y=\sigma_y$, and $\gamma_R=|e|E_z\gamma$ with the above defined $\gamma$.

\subsection{External magnetic field}

To include electron interaction with a perpendicular magnetic field in the monolayer model, we should add to the Hamiltonian a standard Zeeman term (in Eq.~\ref{ham1}): 
\begin{equation}
H_Z=\sum_{i\,\sigma \sigma'\alpha \beta }\gamma_Z\, \mathbf{B}\cdot\mathbf{s}_{\sigma\sigma'}\,\delta_{\alpha \beta}
\,{\hat{c}}^{\dagger }_{i\alpha \sigma }{\hat{c}}_{i\alpha \sigma '},
\end{equation}
with a magnetic field $\mathbf{B}$.
For $\gamma_Z=\frac{g_e\mu_B}{2}$
we arrive at the standard Zeeman energy $\frac{g\mu_B}{2} \mathbf{s} \cdot \mathbf{B}$. 

To address also orbital effects related to magnetic field we apply the so-called Peierls substitution\cite{hofst}. We multiply the hopping matrix, by the additional factor $t_{ij}\rightarrow \tilde{t}_{ij}= t_{ij}\exp\left( \imath\theta_B\right)$ in the Hamiltonian (\ref{ham1}). Now the vector potential enters Eq.~(\ref{ham1}) via the Peierls phase $\theta_B$, calculated as the path integral between neighbor nodes:
\begin{equation}
\theta_B=\frac{e}{\hbar} \int_{\mathbf{r}_i}^{\mathbf{r}_j}\mathbf{A}(\mathbf{r})\cdot d\mathbf{r} .
\end{equation}
$\mathbf{A}$ is a vector potential induced by the $\mathbf{B}$ field.
We use the Landau gauge, with the vector potential $\mathbf{A}(\mathbf{r})=[0,\int dx B_z(x,y),0]^T$ for the perpendicular magnetic field $\mathbf{B}(\mathbf{r}) = [0,0,B_z(\mathbf{r})]^T$. This leads to the phase: 
\begin{equation}
\theta_B=\frac{e}{\hbar}\left(A_y(\mathbf{r}_i)+A_y(\mathbf{r}_j)\right)(y_j-y_i)/2.
\end{equation}

The most important result of applying a magnetic field is a splitting introduced between levels with opposite spin and valley index. Interestingly there are two types of Zeeman splittings: standard, spin type, and Zeeman valley splitting, both presented in Fig.~\ref{fig:6} in Section IV. Each of them possesses other Land\'{e} factor. This will enable us to separately address each transition between four basis states in the spin-valley two-qubit space.

\section{Calculation method}

Let's have a look at the stationary and time-dependent calculation methodology. Firstly, we solve the eigenproblem for the stationary Hamiltonian (\ref{ham1}): $H(\mathbf{r})\boldsymbol{\psi}_m(\mathbf{r}) = E_m \boldsymbol{\psi}_m(\mathbf{r})$, and obtain $M$ eigenstates. For our hexagonal flake of size $N_B=25$ we have $1801\,\mathrm{sites}\,\times 6$ giving $M=10806$ eigenstates. For the Hamiltonian matrix eigenproblem we utilize the fast and efficient FEAST routine\cite{feast}.
The obtained eigenstates are represented by 6-dimensional vectors $\boldsymbol{\psi}_m(\mathbf{r})=(\psi^{\sigma\alpha}_m(\mathbf{r}))^\intercal$, with $\sigma=1,2$ and $\alpha=1,2,3$. They belong to te state space ${\cal H}_2^\mathrm{spin}\otimes{\cal H}_3^\mathrm{oribital}$, with the spin and the 3-dimensional Mo-orbitals space. To identify them, at first we need an electron density calculated as $\rho(\mathbf{r})=|\boldsymbol{\psi}_m(\mathbf{r})|^2$ to determine if given state is localized at the flake edge forming the so-called edge state, or is confined within the quantum-dot.
Secondly, we need to identify the state quantum numbers---valley and spin indices. To do this we utilize similar formulas as in Eqs.~(\ref{eq:valley}) and (\ref{eq:spin}), here adapted to a stationary state $\boldsymbol{\psi}_m(\mathbf{r})$.

During the time-dependent calculations we will be working in the previously found eigenstates base. Therefore the full time-dependent wave function is represented as a linear combination of $N$ basis states $\boldsymbol{\psi}_n$:
\begin{equation}
\label{lin}
\boldsymbol{\Psi}(\mathbf{r},t)=\sum_n c_n(t)\,\boldsymbol{\psi}_n(\mathbf{r})e^{-\frac{\imath}{\hbar}E_n t},
\end{equation}
together with time-dependent amplitudes $c_n(t)$ and phase factors of the corresponding eigenvalues $E_n$. We assume a basis of $N=200 < M$ lowest eigenstates from the conduction band (represented by yellow bullets in the lower inset in Fig.~\ref{fig:5}). The time evolution is governed by the time-dependent Schr\"{o}dinger equation:
\begin{equation}
\label{schr}
\imath\hbar\frac{\partial}{\partial t} \boldsymbol{\Psi}(\mathbf{r},t) = H(\mathbf{r},t) \boldsymbol{\Psi}(\mathbf{r},t),
\end{equation}
with the time-dependent Hamiltonian being a sum of the stationary part (Eq.~\ref{ham1}) and a time-dependent contribution to both the potential energy and the Rashba coupling:
\begin{equation}
\label{var}
H(\mathbf{r},t)=H(\mathbf{r})+\delta\varphi(\mathbf{r},t)+\delta H_R(\mathbf{r},t).
\end{equation}
The full time-dependent potential energy $\varphi(\mathbf{r},t)=\varphi(\mathbf{r})+\delta\varphi(\mathbf{r},t)$ contains variable part 
$\delta\varphi(\mathbf{r},t)$, generated by modulation of the gate voltages. Whole is calculated as $\varphi(\mathbf{r},t)=-|e|\phi(\mathbf{r},t)$, with the potential $\phi(\mathbf{r},t)$ obtained by solving the Poisson equation for the variable density $\rho(\mathbf{r},t)$ at every time step. Note that the charge density originates from the actual wave-function, thus the Schr\"{o}dinger and Poisson equations are solved in a self-consistent way. Similarly, the time-dependent part of the Rashba coupling $\delta H_R(\mathbf{r},t)$ is induced by the variable part of the electric field: $E_z(\mathbf{r},t)=E(\mathbf{r})+\delta E_z(\mathbf{r},t)$. That is, $E(\mathbf{r})$ induces 
(\ref{rash}), while $\delta E_z(\mathbf{r},t)$ enters $\delta H_R(\mathbf{r},t)$ in (\ref{var}) with the same formula.  

Insertion of (\ref{lin}) to the Schr\"{o}dinger equation (\ref{schr}) gives a system of equations for time-derivatives of the expansion coefficients at subsequent moments of time:
\begin{equation}
\label{timeq}
\dot{c}_m(t)=-\frac{\imath}{\hbar}\sum_n c_n(t)\, \delta_{mn}(t)\, e^{\frac{\imath}{\hbar}(E_m-E_n)t}.
\end{equation}
The actual matrix elements $\delta_{mn}(t)=\langle \boldsymbol{\psi}_m| \delta\varphi(\mathbf{r},t) + \delta H_R(\mathbf{r},t)|\boldsymbol{\psi}_n\rangle$ need to be calculated at every time step due to changes in the potential and the electric field. Then, by using it, we solve the system (\ref{timeq}) iteratively using a predictor-corrector method, with explicit ``leapfrog'' and implicit Crank-Nicolson scheme, obtaining the next time step of the system evolution. 

For the electron wave function $\boldsymbol{\Psi}(\mathbf{r},t)$ we calculate the Fourier transform:
\begin{equation}
\tilde{\boldsymbol{\Psi}}(\mathbf{k},t)=\int_{F}\!d^2r\,\boldsymbol{\Psi}(\mathbf{r},t) e^{-\imath\mathbf{k}\mathbf{r}},
\end{equation} 
on the flake surface area $F$, with 2D-wave vector $\mathbf{k}\equiv(k_x,k_y)$. The Fourier transform naturally has periodic structure in the reciprocal space, therefore we can limit the $k$-area to $\tilde{F}$: $k_{x,y}\in\left[-\frac{2\pi}{a},\frac{2\pi}{a}\right]$, encompassing the (first) Brillouin Zone (BZ). Knowing $\tilde{\boldsymbol{\Psi}}(\mathbf{k},t)$ we can calculate density in the reciprocal space expressed as: $\tilde{\rho}(\mathbf{k},t)=|\tilde{\boldsymbol{\Psi}}(\mathbf{k},t)|^2$.
The $k$-density calculated for the $|K\!\!\downarrow\rangle$ state from Fig.~\ref{fig:5} is presented in Fig.~\ref{fig:10}($0$). We also mark the BZ along with points of high symmetry: $\Gamma$ in the center, two types of $K$($K’$) at the corners of the hexagonal zone and $M$ on the edges of the hexagon. The coordinates of the high-symmetry points are:
$\Gamma=(0,0)$, one of $K=\frac{\pi}{a}(\frac{4}{3},0)$ 
and one of $M=\frac{\pi}{a}(1,\frac{1}{\sqrt{3}})$, 
with the lattice constant $a = 0.319$~nm. We can clearly see in Fig.~\ref{fig:10}($0$) that density peaks are localized in the neighborhood of the $K$ points (while not next to $K'$), confirming that in the $|K\!\!\downarrow\rangle$ state exactly $K$ valley is occupied.

Now, the valley index $\mathcal{K}$ is calculated as:
\begin{equation}
\label{eq:valley}
\mathcal{K}(t)=\frac{3a}{4\pi}\int_{\tilde{F}_{1/3}}\!d^2k\,\tilde{\rho}(\mathbf{k},t)k_x
\end{equation}
on the reciprocal space area $\tilde{F}_{1/3}$  defined as two opposite $\pi/3$ sectors (within $\tilde{F}$ area) encompassing exactly one $K$ point and one $K'$ point, i.e. $|\vartheta|\le\frac{\pi}{6}\cup |\vartheta|\ge\frac{5\pi}{6}$, with azimuthal angle $\vartheta=\mathrm{atan2}(k_x,k_y)$. Because $K$($K'$) point in $\tilde{F}_{1/3}$ has coordinates $1$($-1$)$\times(\frac{4\pi}{3a},0)$, the valley index $\mathcal{K}\in\left[-1,1\right]$. $\mathcal{K}=1$ represents the $K$ valley, whereas $\mathcal{K}=-1$ the $K'$ valley. For example the state $|K\!\!\downarrow\rangle$ is thus represented by $|\mathcal{K},s\rangle=|1,\downarrow\rangle$ ket.

The electron spin value $s$ is calculated as the expectation value of the Pauli $z$-matrix $s^z=\sigma_z$:
\begin{equation}
\label{eq:spin}
s(t)=\int_{F}\!d^2r\,\boldsymbol{\Psi}^\dag(\mathbf{r},t)\, \sigma_z\otimes\mathbf{1}_3\,\boldsymbol{\Psi}(\mathbf{r},t),
\end{equation}
also integrated on the flake area $F$ for the actual wave vector 
$\boldsymbol{\Psi}(\mathbf{r},t)$. Operation $\otimes\,\mathbf{1}_3$ means that during the spin calculations we trace out over the orbitals subspace.

\section{Electrostatic quantum dot}

\begin{figure}[b]
	\center
	\includegraphics[width=7cm]{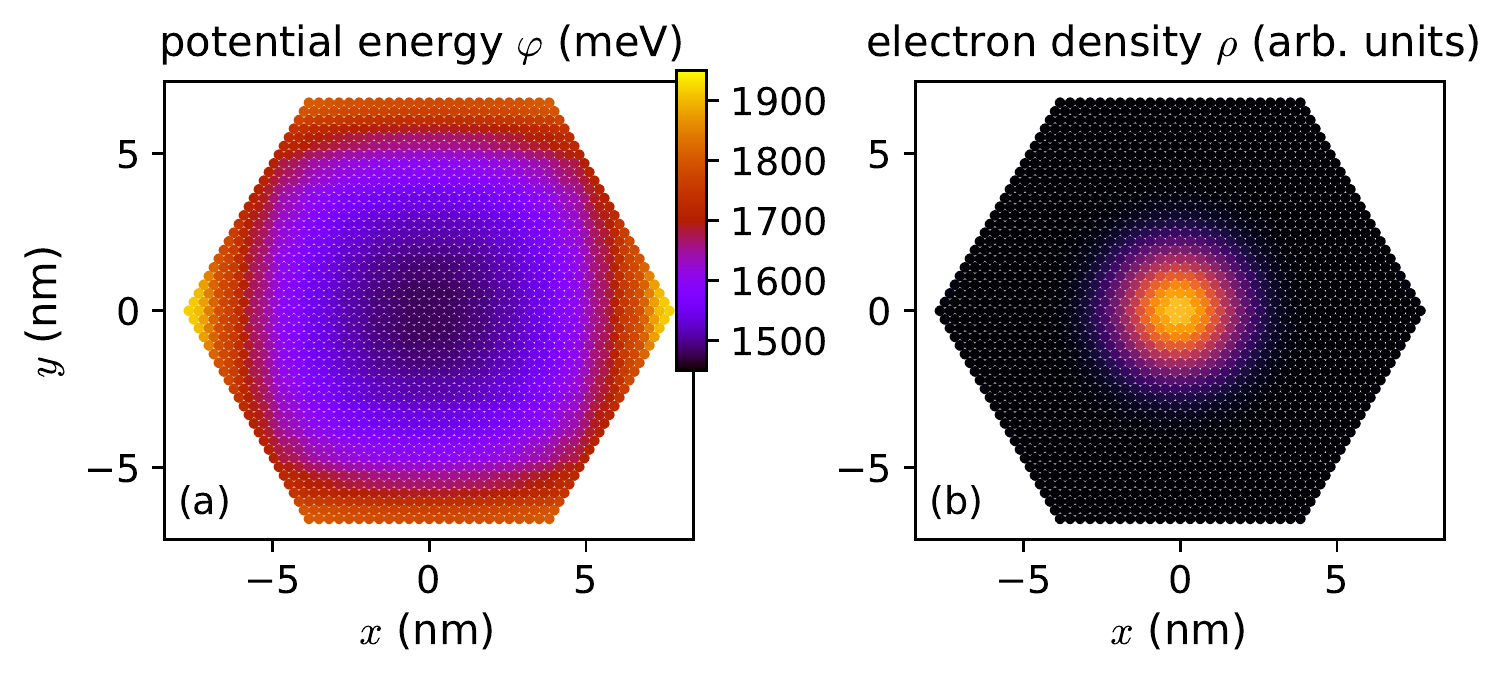}
	\caption{\label{fig:4} (a) The confinement potential energy, as in Fig.~\ref{fig:2}, induced in the MoS$_2$ flake area forming QD in which we confine a single electron. (b) The density of the electron lowest state in the CB minimum.}
\end{figure}

\begin{figure}[t]
	\center
	\includegraphics[width=8.2cm]{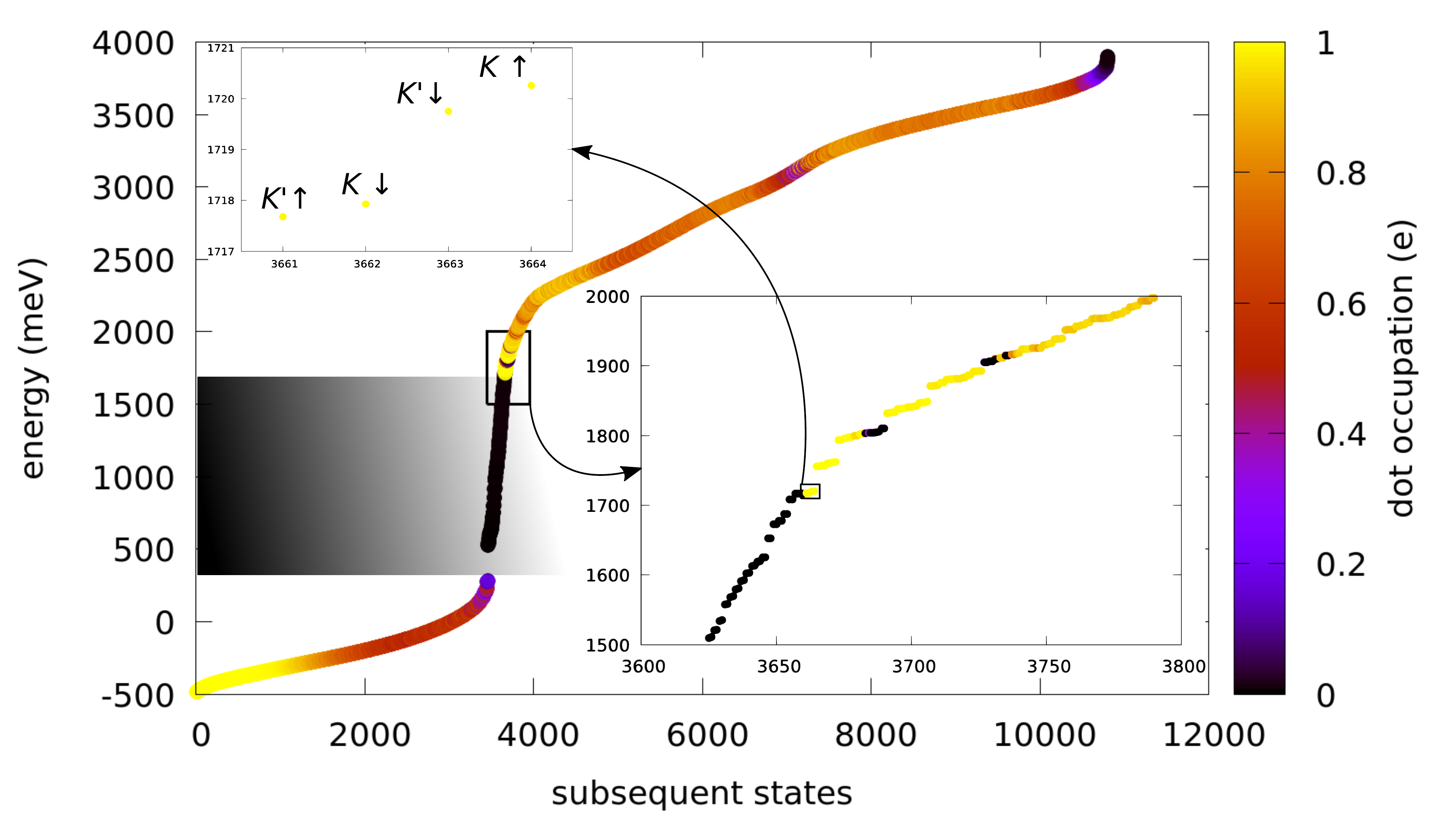}
	\caption{\label{fig:5} Subsequent eigenenergies of the flake Hamiltonian (\ref{ham1}) for an electron confined in the created quantum dot, marked as color bullets. The bullet-colors describe the QD occupation, with yellow states for carriers strongly confined and localized at the flake center, while black bullets for the edge states with density at the flake border. The latter levels are forbidden for a carrier confined within the dot and define the energy bandgap.}
\end{figure}

By applying voltages $V_{1..4}=1500$~mV to all of the gates we form the QD potential energy in the flake area, presented in Fig.~\ref{fig:4}(a). Calculated electronic eigenstates of the Hamiltonian (\ref{ham1}) for the entire flake lattice, forms a ladder in Fig.~\ref{fig:5} representing subsequent $M=10806$ eigenstates $\boldsymbol{\psi}_m(\mathbf{r})$. The bullets color is used to mark the dot occupation, namely the brighter the color is, the electron is more localized in the flake center. E.g. the yellow states are strongly confined, while black color marks the edge states with density localized on the flake border. These states are inaccessible to the electron confined in the QD, forming a forbidden energy range, namely, a bandgap. The bandgap divides the QD eigenstates into conduction and valence bands. Lets now zoom into CB minimum. The states therein are presented in insets from Fig.~\ref{fig:5}.

The first four states form two doublets $\{|\!-\!1,\uparrow\rangle,|1,\downarrow\rangle\}$ and $\{|\!-\!1,\downarrow\rangle,|1,\uparrow\rangle\}$, spin-orbit split (see the upper inset in Fig.~\ref{fig:5}). Their electron density is presented in Fig.~\ref{fig:4}(b).
It turns out, that (states from) both the bottom of the conduction band and the top of the valence band are located at the points $K$ and $K’$ (we have a direct band gap here), not at the $\Gamma$ point. These bands form two non-equivalent valleys $K$ and $K’$ which can be occupied by qubit carriers. Subspace spanned by the first four states consists of exactly one valley and one spin two-level system, forming together a 4-dimensional Hilbert space ${\cal H}_2^\mathrm{valley}\otimes{\cal H}_2^\mathrm{spin}$ of spin-valley two-qubit states $|\mathcal{K},s\rangle$. 

\subsection{Two-qubit subspace}

\begin{figure}[t]
	\center
	\includegraphics[width=7cm]{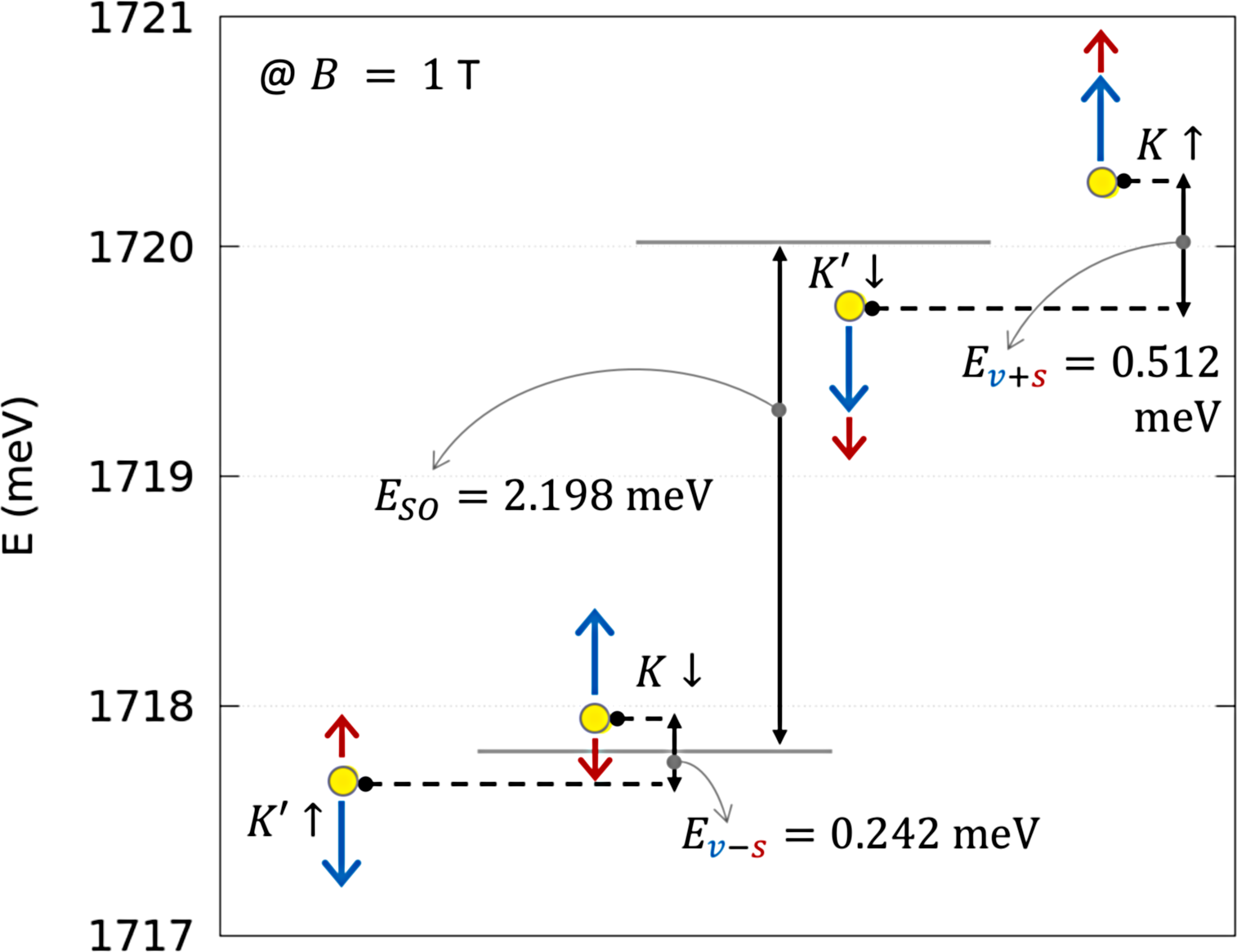}
	\caption{\label{fig:6} The four lowest CB states (yellow bullets) forming the two-qubit spin-valley subspace. Applied magnetic field $B_z=1$~T induces spin (represented by red arrows) and valley (blue arrows) Zeeman splitting, enabling separate addressing of all six transitions within the two-qubit subspace.}
\end{figure}

If we add an external magnetic field, degeneracy in both pairs is lifted, as presented in Fig.~\ref{fig:6} for $B_z=1$~T. 
Calculated splitting, here for $1$~T (for $2$~T resulting factors are the same), between first pair: $\mathcal{E}_{K'\uparrow}=\frac{1}{2}(g_s-g_v)\mu_{B}B$ (split down) and $\mathcal{E}_{K\downarrow}=\frac{1}{2}(-g_s+g_v)\mu_{B}B$ (split up) is $242$~$\mu$eV, which is in agreement with differences $238$~$\mu$eV between opposite spin states (with different valleys) in the conduction band minimum for a larger dot [\onlinecite{dias}]. The difference of $242$~$\mu$eV leads to $g_v-g_s\simeq4.18$. 
For the second pair
$\mathcal{E}_{K'\downarrow}=\frac{1}{2}(-g_s-g_v)\mu_{B}B$ (split down) and $\mathcal{E}_{K\uparrow}=\frac{1}{2}(g_s+g_v)\mu_{B}B$ (split up) we have $512$~$\mu$eV, thus $g_v+g_s\simeq8.84$. Therefore, obtained effective valley and spin g-factors are: $g_v=6.51$, similar to DFT calculations giving the value 7.14 [\onlinecite{prx}, and errata: \onlinecite{prxe}]. While the spin splitting factor $g_s=2.33$, with agreement with the DFT calculations [\onlinecite{prx}] and experimental result [\onlinecite{gs}], both giving value about $2.2$.

If we now take into account both spin and valley splitting, it turns out that the higher states pair is more split than the lower one, as presented in Fig.~\ref{fig:6}. This results from emergence of additional valley Zeeman splitting, for which $K$ levels bend upwards and for $K'$ downwards (blue arrows). Similarly, spin-down level bends down, while spin-up bends up (red arrows). Therefore, for the higher levels pair the splittings will add, while for the lower one they will subtract. Thanks to such a form of level splittings, all of the transitions between them (6 in total) can be separately addressed.

\subsection{Intervalley coupling}
Let's now examine the intervalley coupling strength and its origin.  
The doubled intervalley coupling $2\Lambda$ can be calculated from the difference between the ground state and the 1st excited state in CB with no spin-orbit interaction\cite{2lambda}. Further, we assume that the intervalley coupling is: $\Lambda(t)=\Lambda_0+\frac{\Lambda_1}{2}\cos(\omega t)$, with a modulation amplitude $\Lambda_1$.
\begin{figure}[h]
	\center
	\includegraphics[width=8cm]{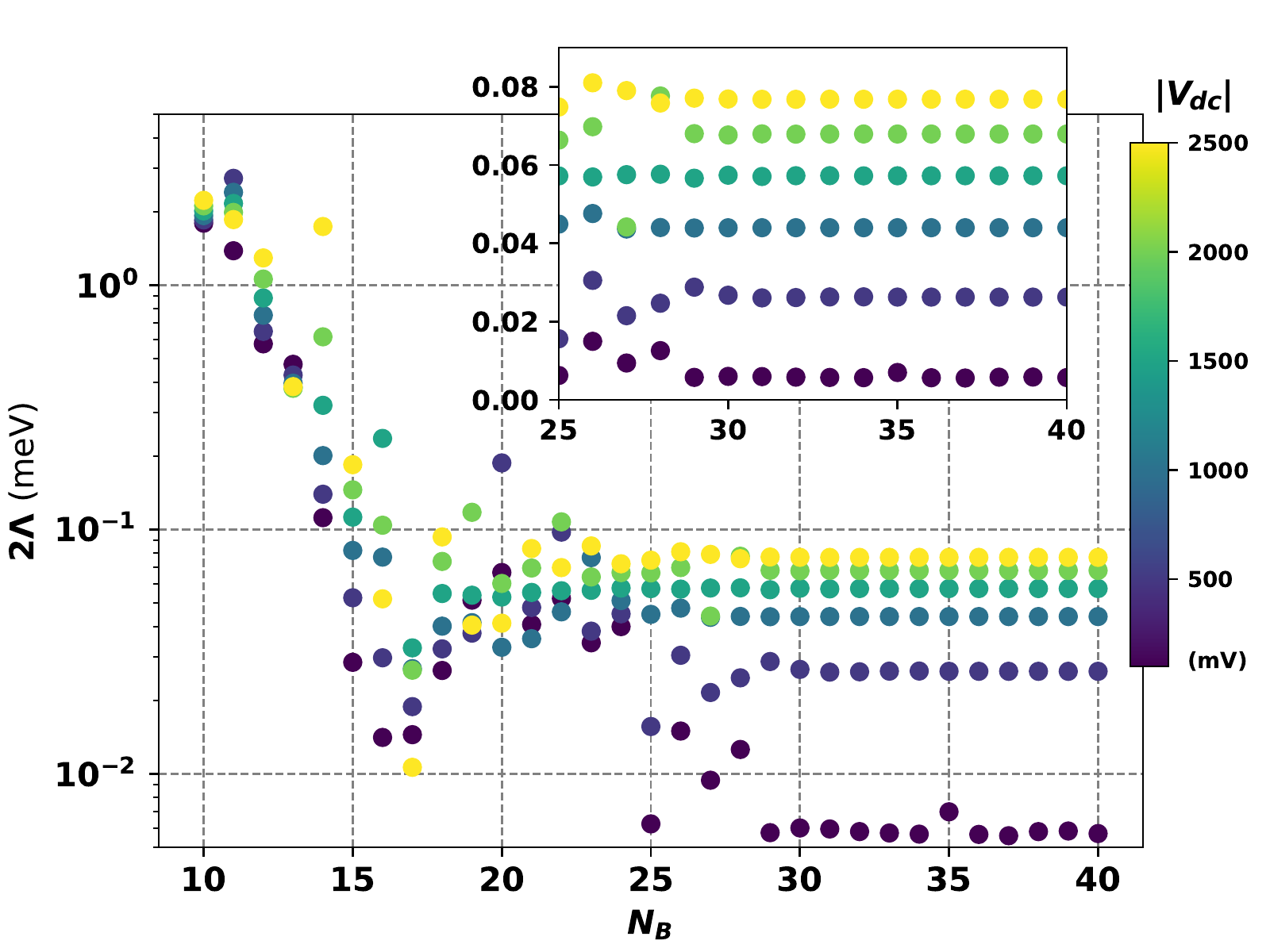}
	\caption{\label{fig:7} Intervalley coupling $2\Lambda$ as a function of the flake size $N_B$ and the applied voltage. Rising the voltage amplitude $V_{dc}$ increases the confinement depth, which results in stronger intervalley coupling.}
\end{figure}

The structure of electronic states confined within a nanoflake and mostly presence of edge states, that cross the gap, depends on the flake edge type\cite{edge0,edge1,edge2}. Zigzag edges do not mix valleys, but supports edge states. On the other hand, gap states are missing for an armchair edge type, which mixes valleys and induces transitions in graphene-like structures\cite{edge3}. Same edge-dependent valley mixing was proven for MoS$_2$ nanoribbons\cite{edge4}. Here, to skip the edge influence, we take a flake with a zigzag edge, and induce confinement strong enough to decouple the electron from the flake edge. 

To check completely the edge influence on the intervalley mixing, we calculate $2\Lambda$ as a function of the flake size $N_B$ and the confinement depth, controlled by $V_{dc}$. The results with no confinement are presented in Fig.~\ref{fig:7}, by dark-blue bullets. While there is significant coupling for very small flakes with $N_B\sim10$, it suddenly decreases with the flake size, reaching two orders smaller value for $N_B\sim15$. Then the coupling slightly increase, but for larger flakes with $N_B>25$ it generally does not exceed several $\mu$eV. However if we add the confinement potential by applying $V_{dc}$, the intervalley coupling increases with the potential depth, which is clearly visible for $N_B>25$. Subsequent voltage values, marked by the brighter bullets, are $V_{dc}$ = $-500, -1000, -1500, -2000, -2500$~mV. It turns out that the approximate relation between the coupling and the applied voltage in this range of the flake sizes is $2\Lambda \sim \sqrt{V}$. It is similar to the relation between eigenenergies and potential of a harmonic oscillator. 
Now the coupling is purely confinement-dependent and does not depend on the flake size. This is because the confined electron is decoupled from the edge and its valley mixing is controlled electrically via the confinement potential.

A 10-nm-scale lithographic process required in in Fig.~\ref{fig:1} can be difficult to achieve. Our calculations show that scaling up the structure (with preserved gate voltages) will decrease the intervalley coupling amplitude as $1/L$, where $L$ is the gate width. This means that after magnifying gates layout by an order of magnitude the coupling still keeps reasonable values. However, such scaling proportionally extends the intervalley transition time.

\section{Spin-valley two qubit system}
\begin{figure}[t]
	\center
	\includegraphics[width=8cm]{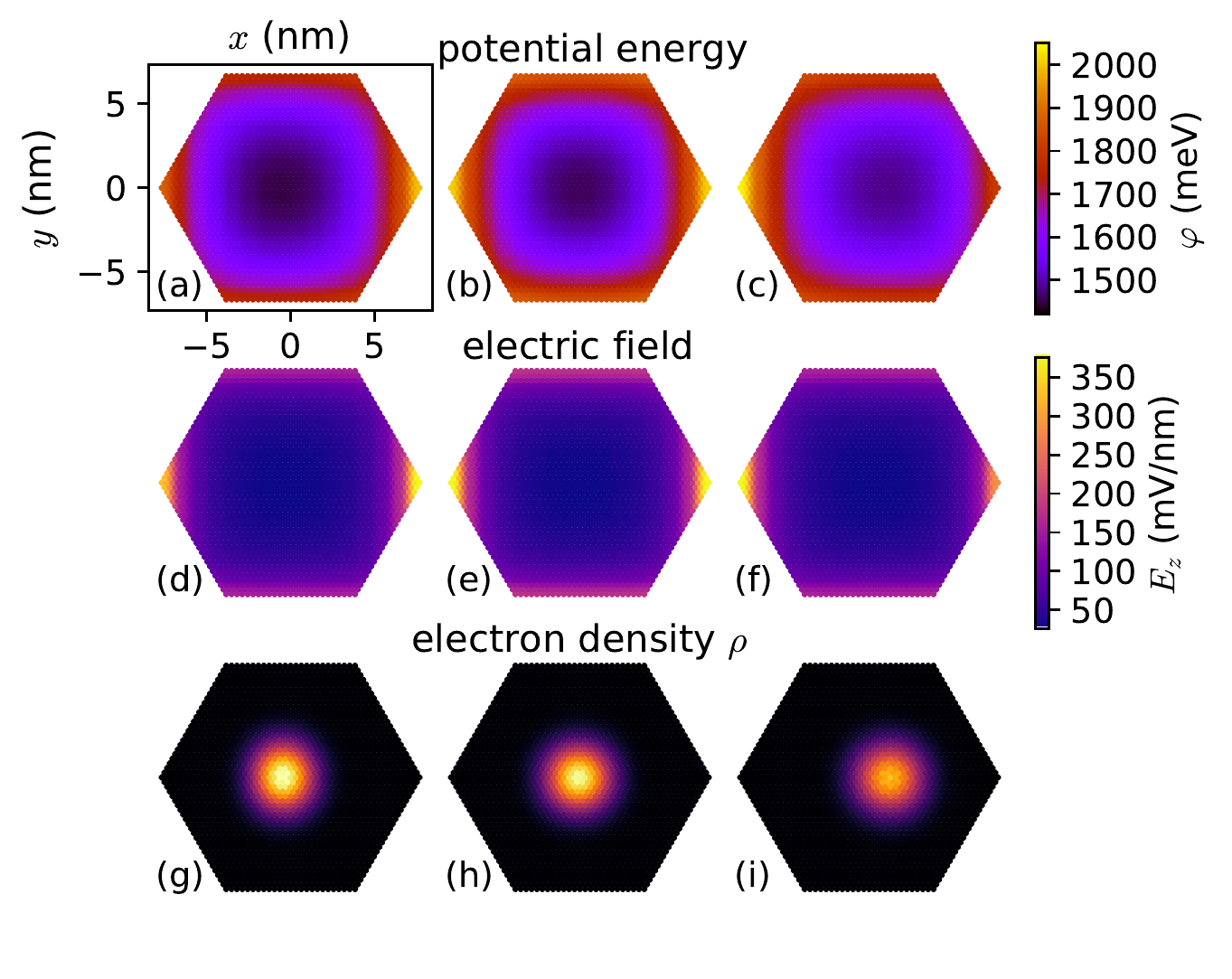}
	\caption{\label{fig:8} The confinement modulation: (a-c) the gate defined electrostatic potential $\varphi$ within the flake at three instants: $t=\frac{\pi}{2\omega}$ (left column), $t=0$ (middle), and $t=\frac{3\pi}{2\omega}$ (right); (d-f) the perpendicular electric field $E_z$ at the same instants; (g-i) the confined electron density is squeezed and moved by the corresponding potential modulation.}
\end{figure}

Let's now switch to the time-dependent calculations and examine the process of inducing transitions within the defined spin-valley subspace.
By applying additional oscillating voltage to the single control gate $G_{1}$: $V_1(t)=V_{dc}+V_{ac}\sin(\omega t)$, $V_{ac}=-100$~mV together with  $V_2=V_3=V_4=V_{dc}=-2500$~mV,  we modulate the confinement potential in a way that the dot minimum oscillates back and forth in the $x$-direction. The potential energy landscape at oscillations start, i.e. at $t=0$, is presented in Fig.~\ref{fig:8}(b). While its form at the maximum left and right displacement from the center position, e.g. at $t=\frac{\pi}{2\omega}$ and $t=\frac{3\pi}{2\omega}$, is presented in Figs.~\ref{fig:8}(a) and \ref{fig:8}(c) respectively. Additionally to inducing oscillatory movement of the dot position, the potential shape is modulated and becomes narrower at the maximum shift to the left---see Fig.~\ref{fig:8}(a), while shallower at the maximum right---Fig.~\ref{fig:8}(c). This enforces oscillatory squeezing of the electron state density, as seen in Fig.~\ref{fig:8}(g-i).

\subsection{Spin and valley transitions}
\begin{figure}[t]
	\center
	\includegraphics[width=8cm]{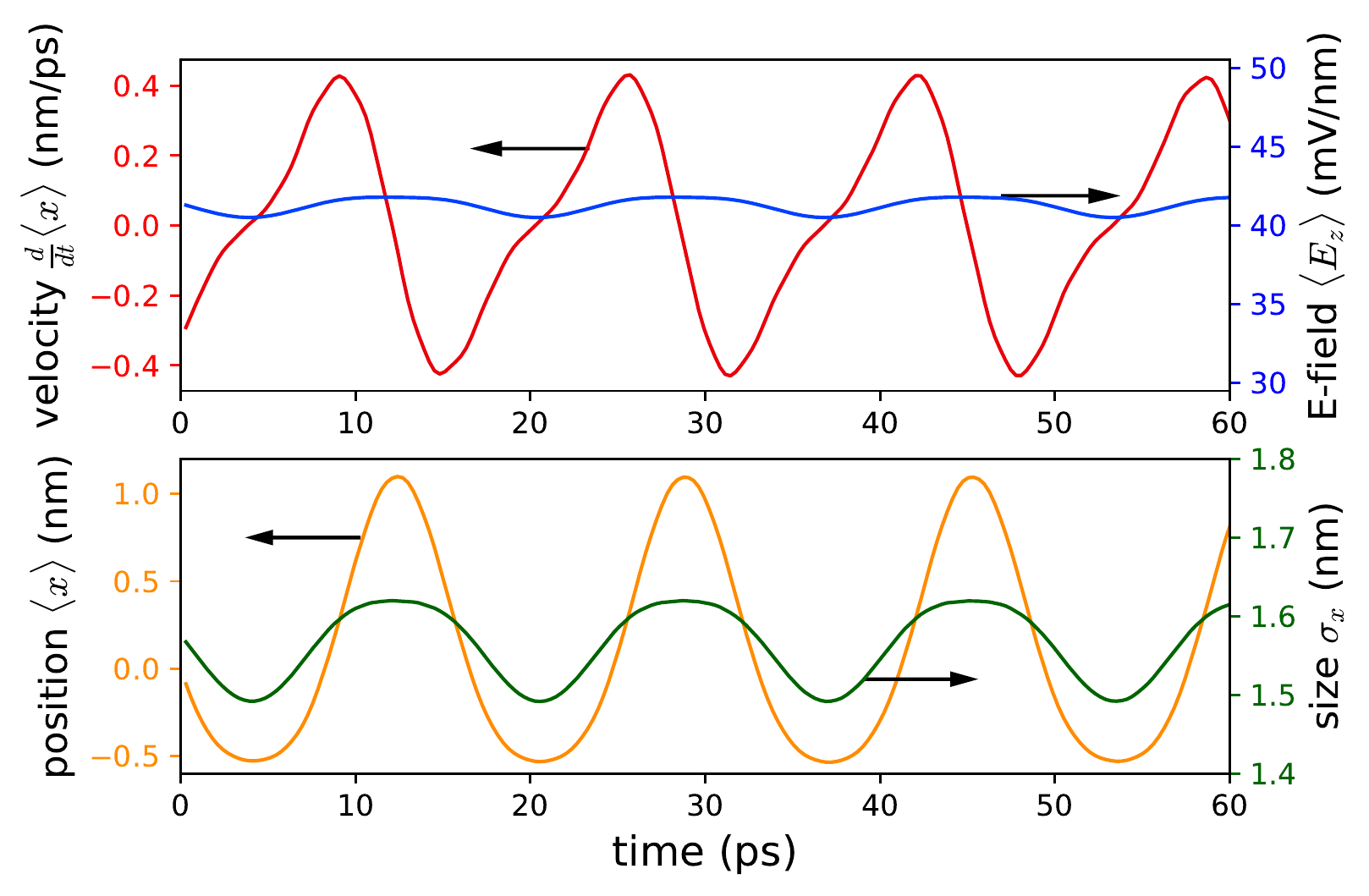}
	\caption{\label{fig:9} The gate $G_1$ voltage modulation $V_{ac}=-100$~mV provides two simultaneous effects: oscillatory shifting of the confinement minimum moves the electron position (orange curve) and thus generates the electron momentum (red curve), which together with the Rashba SOI, introduced by a perpendicular electric field (blue curve), induces the electron spin transitions. Simultaneous potential narrowing squeezes the electron density, modulates the electron packet size (green curve) and provides the intervalley transitions.}
\end{figure}
The confinement potential modulation introduces two effects to the system. Firstly, the voltage modulation moves the electron confined in the QD potential in an oscillatory way. The electron position oscillations causes that its momentum also oscillates. The velocity defined as the time derivative of the electron expectation position $\frac{d}{dt}\langle x\rangle$ is presented in Fig.~\ref{fig:9} as the red curve, while the electron position as the orange one.
Together with the present perpendicular electric field $\langle E_z\rangle$ felt by the electron (blue curve in Fig.~\ref{fig:9}) which induces the Rashba SOI, it creates spin-orbit mediated electron spin resonance transitions. The mean electric field $\langle E_z\rangle$ is almost constant during oscillations, which is related to fact that the perpendicular electric field component $E_z(x,y)$ is mostly uniformly distributed on the flake, as seen in Fig.~\ref{fig:8}(d-f).

Secondly, oscillatory shallowing of the confinement potential leads to electron packet squeezing, visible as oscillations of the electron packet size $\sigma$ in Fig.~\ref{fig:9} (green curve) and causes intervalley coupling changes. Resonant modulation of the intervalley coupling generates gradual transitions of the electron between the different valley states\cite{mymos2}.

\begin{figure}[b]
	\center
	\includegraphics[width=8.2cm]{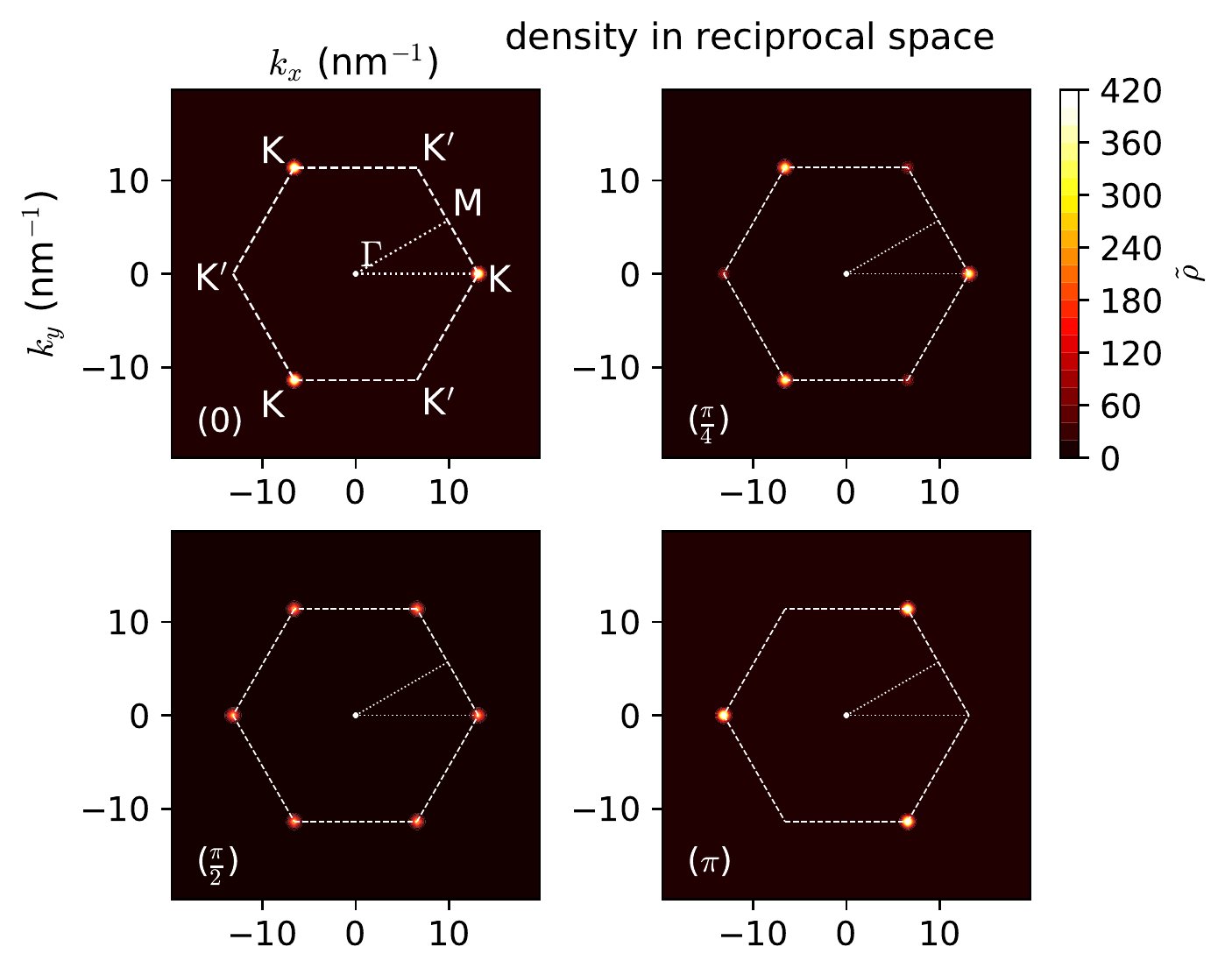}
	\caption{\label{fig:10} Successive stages of the intervalley transitions visible as gradual flow of the electron from the $K$ to $K'$ valley. The electron density in the reciprocal space $\tilde{\rho}(k_x,k_y)$ initially localized in the $K$ valley in the BZ ($0$), starts to flow into $K'$ point ($\frac{\pi}{4}$), reaches equal occupation ($\frac{\pi}{2}$), and finally entirely occupies the $K'$ valley ($\pi$).}
\end{figure}

\begin{figure}[b]
	\center
	\includegraphics[width=8cm]{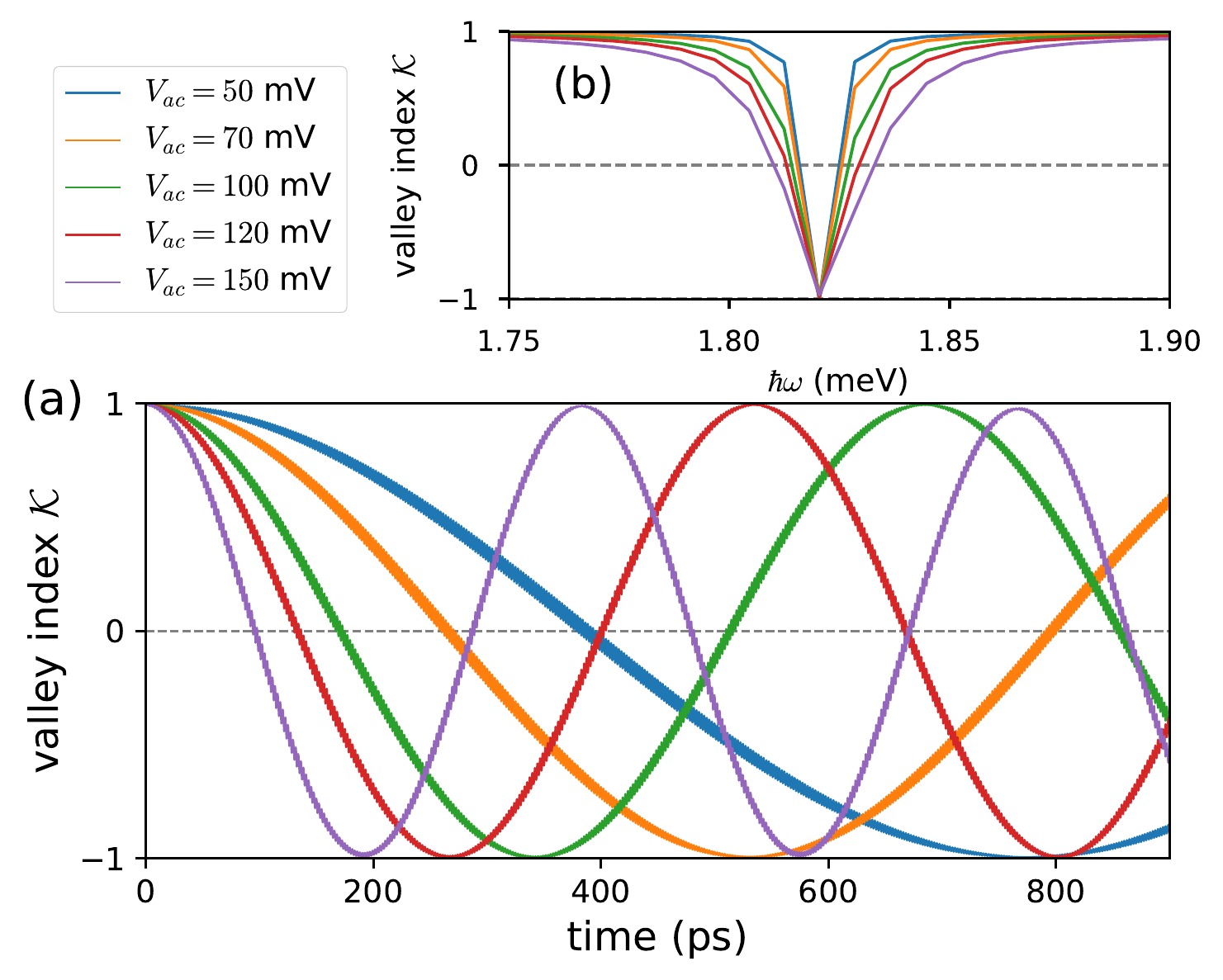}
	\caption{\label{fig:12} The intervalley transitions for different driving voltage $V_{ac}$ amplitudes. The transition frequency $\Omega=\frac{2\pi}{T}$ (with $T$ being the transition period) and the resonance peak width are proportional to the driving amplitude, which is characteristic of the Rabi oscillations.}
\end{figure}

Subsequent stages of a transition between the different valleys in reciprocal space are presented in Fig.~\ref{fig:10}.
Initially, the electron density in the reciprocal space $\tilde{\rho}(k_x,k_y)$ is localized in the $K$ point vicinity within the BZ, as showed in Fig.~\ref{fig:10}($0$). The voltage pumping process with the resonant frequency $\omega$, tuned to the energy spacing $\hbar\omega$ between the states $|1,\uparrow\rangle$ and $|\!-\!1,\uparrow\rangle$ (or $|1,\downarrow\rangle$ and $|\!-\!1,\downarrow\rangle$), leads to a gradual change of occupation to $K'$ valley, with density flow between valleys visible in the subsequent stages---Fig.~\ref{fig:10}($\frac{\pi}{4}$) and ($\frac{\pi}{2}$). After time $t=345$~ps the electron occupies the $K'$ valley entirely (Fig.~\ref{fig:10}($\pi$)) and the intervalley transition is completed. The whole process is presented in Fig.~\ref{fig:12}, where the green curve represents the valley index $\mathcal{K}$ evolution during the entire transition.

\begin{figure}[b]
	\center
	\includegraphics[width=8cm]{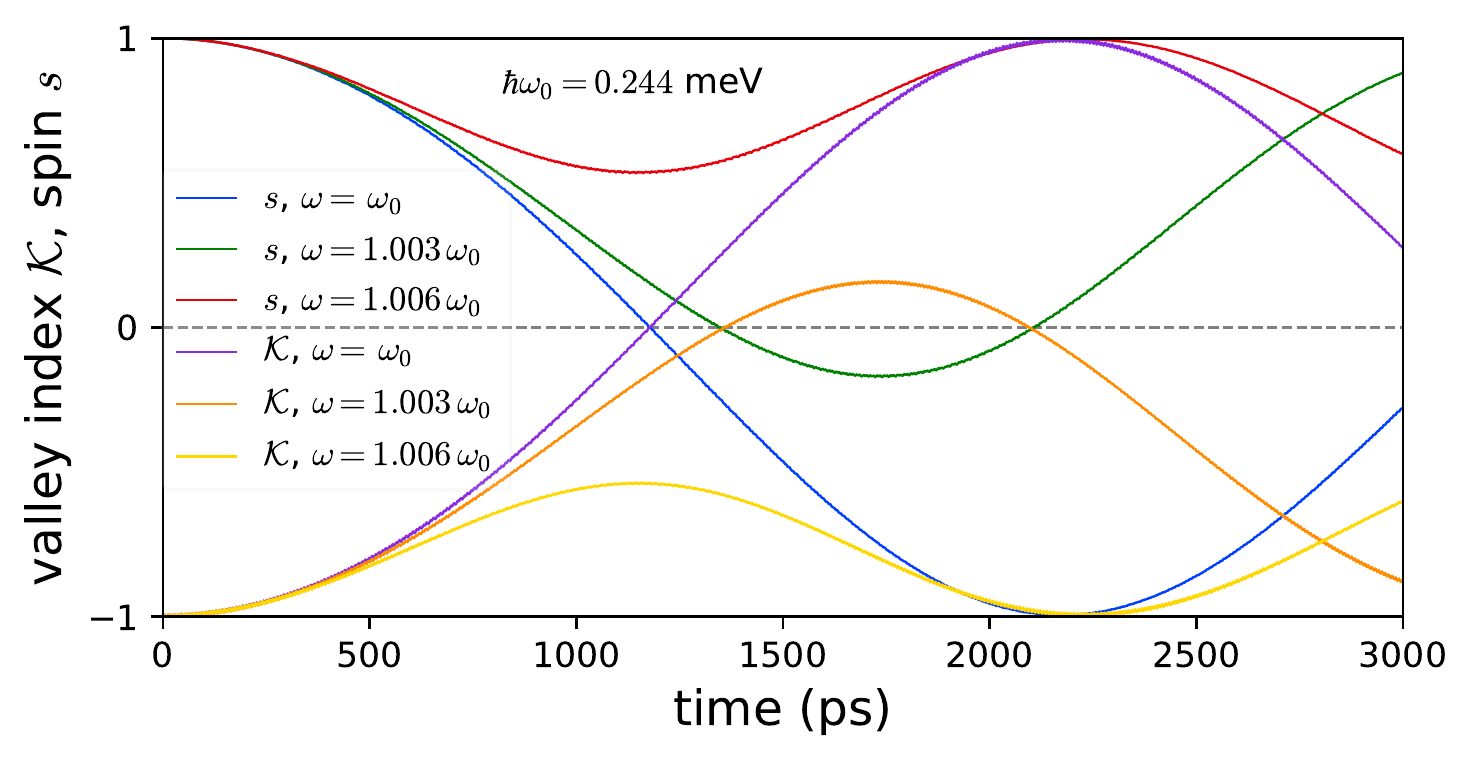}
	\caption{\label{fig:11} The spin-valley SWAP operation exhibiting Rabi oscillations form: blue and violet curves present spin and valley index swaps at resonance leading to full transitions. Whereas beyond the resonance we observe incomplete flips: the green and orange curve pair shows transitions closer, while the red and yellow further from the resonance.}
\end{figure}

Besides inter- spin and valley transitions we can simultaneously obtain spin and valley manipulation leading to inter spin-valley transitions or simply the spin-valley SWAP. Similarly here transitions are resonant and we need to tune the modulation frequency $\omega$ to the energy spacing between $|\!-\!1,\uparrow\rangle$ and $|1,\downarrow\rangle$ (or $|\!-\!1,\downarrow\rangle$ and $|1,\uparrow\rangle$). During voltage oscillations both the Rashba SOI mediated spin transitions and the intervalley coupling modulation effects are enabled, thus allowing for simultaneous spin and valley flipping. Both effect are needed: turning off the Rashba SOI, by setting $E_z(x,y)=0$, turns off the spin-valley SWAP. Simulation results are presented in Fig.~\ref{fig:11} with the resonance frequency $\hbar\omega_0=0.244$~meV and the driving voltage $V_{ac}=100$~mV. We observe here simultaneous spin (blue curve) and valley index (violet curve) flips.
These transitions are obviously of the Rabi oscillations type. If we diverge from the resonance, the maximum (minimum) value of the valley (spin) index falls down rapidly, entering the region of incomplete transitions. In Fig.~\ref{fig:11}, green and orange (red and yellow) curves pair present spin and valley index courses for a driving frequency $\omega=1.003\,\omega_0$ ($\omega=1.006\,\omega_0$) beyond the resonance.

In opposite to spin-valley SWAP, the intervalley transitions are not mediated by the spin-orbit coupling, and are unaffected even if we eliminate the electric field in the simulations, simply by setting $E_z(x,y)=0$ in (Eq.~\ref{rash}). Moreover, the transition time depends on the amplitude of the intervalley coupling modulation. In Fig.~\ref{fig:12}(a) are presented transitions between both valleys, starting from $K$ valley with the index $\mathcal{K}=1$. 
The transition period $T$ deceases as the voltage modulation amplitude $V_{ac}$ increases. Indeed, for presented in Fig.~\ref{fig:12} amplitude $V_{ac}$ ranges ($50$--$150$~mV), oscillations frequency $\Omega$ turns out to be approximately proportional to $V_{ac}$, and thus to the intervalley coupling modulation amplitude $\Lambda_1$ (we assume here that for such a small voltage modulation range the intervalley coupling $2\Lambda$ responses linearly---cf. Fig.~\ref{fig:7}). This is typical for the Rabi oscillations, where near the resonance the Rabi frequency $\Omega=\sqrt{(\omega-\omega_0)^2+(\Lambda_1/\hbar)^2}$ depends linearly on the driving amplitude $\Lambda_1$ of the intervalley coupling oscillations. In case of resonance $\Omega=\Lambda_1/\hbar$.

If we calculate the minimum value reached by the $\mathcal{K}$ index for out-of-resonance transitions, we obtain resonance curves presented in Fig.~\ref{fig:12}(b). The full width at half maximum (FWHM) parameter characterizing resonance curves in case of the Rabi oscillations corresponds to the transition duration. The resonance curve (for driving $\frac{\Lambda_1}{2}\cos(\omega t)$) has the form $\frac{(\Lambda_1/\hbar)^2}{(\omega-\omega_0)^2+(\Lambda_1/\hbar)^2}$, which gives FWHM equal $2\Lambda_1$. This agrees with our calculations. E.g. for the green curve, i.e. $V_{ac}=100$~mV, transition time (period) $T=2\pi/ \Omega$ is $680$~ps, which corresponds to $\Lambda_1=h/T\simeq0.006$~meV and agrees with FWHM equal $0.012$~meV. In comparison, for the violet curve FWHM is just over two times wider than for orange one.

\subsection{Two-qubit gates}

\begin{figure}[h]
	\center
	\includegraphics[width=8.2cm]{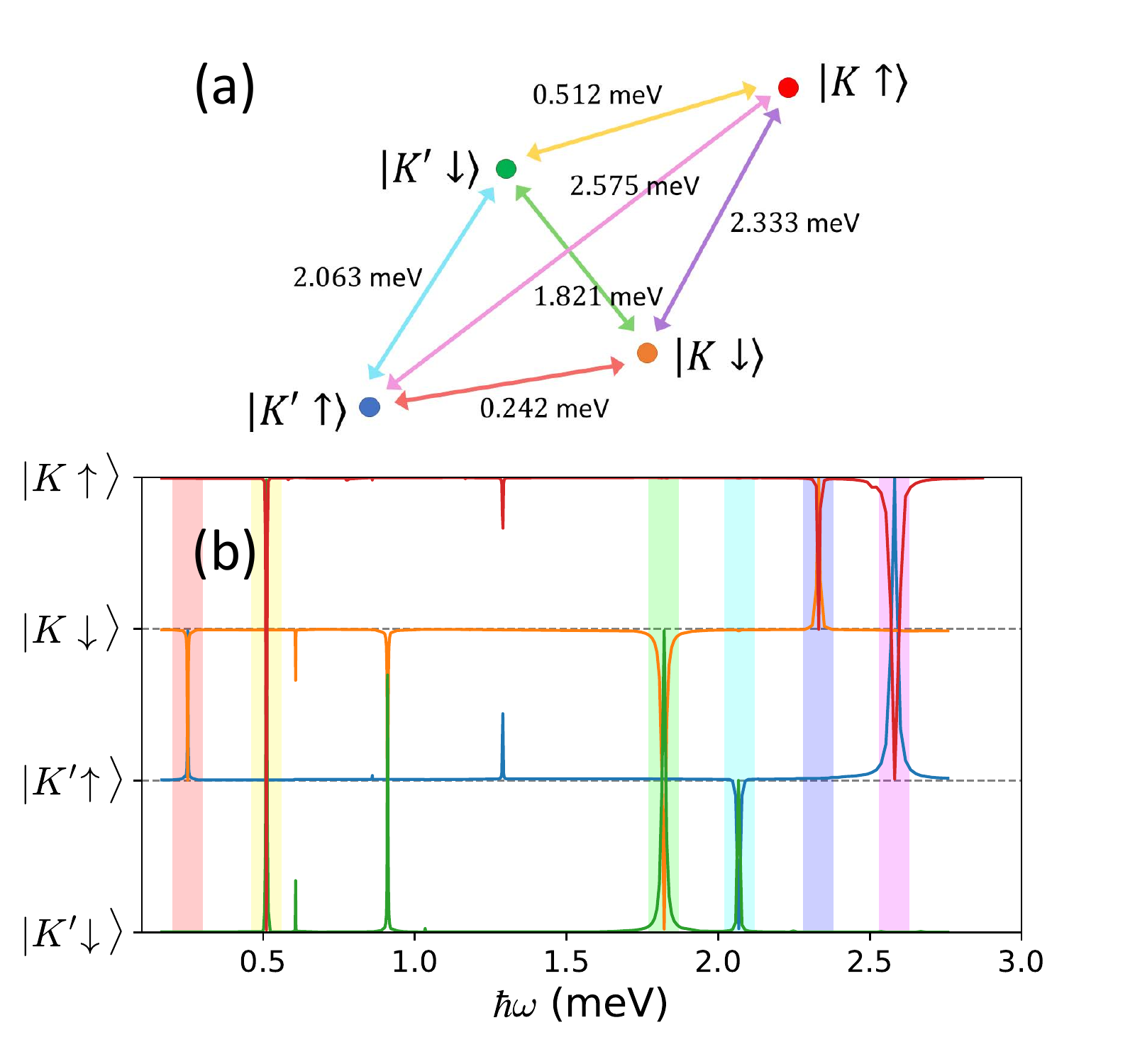}
	\caption{\label{fig:13} Tunning of the driving frequency enables to reach all the transitions within spin-valley two-qubit subspace.}
\end{figure}
In Fig.~\ref{fig:13}(a) there are presented four basis states spanning the two-qubit subspace together with all six transitions---each with a different resonant frequency. The voltage modulations induce electron momentum oscillations and intervalley coupling amplitude modulations which enables us to obtain spin operations (blue and violet arrow pair), intervalley transitions (green and purple), or spin-valley swapping  (red and yellow). This twofold control of the electron state allows to fully operate within the defined spin-valley two-qubit subspace. Simply, by tuning the modulation frequency we can select and switch-on desired transition. 

If we apply a proper magnetic field value (we assume $B_z=1$~T), the Zeeman splitting together with the  spin-orbit induced splitting result in different frequencies among transitions (and allows to separately address each of them). If we now sweep the driving frequency over a range covering all the six transitions, assuming that the system can be initially in four different basis states we observe in Fig.~\ref{fig:13}(b) all of transitions at their own frequencies. They are highlighted by colors corresponding to colors of the arrows from the scheme in Fig.~\ref{fig:13}(a). Interestingly we also observe some minor fractional resonances for lower frequencies $\omega/2$. Fortunately, they do not overlap with other peaks and transitions are not disturbed by each other. The driving amplitude applied in presented simulations is $V_{ac}=100$~mV, with an exception for SWAPs where $V_{ac}=150$~mV.

Let us now translate obtained transitions to the language of qubit operations. Starting from the blue transition from Fig.~\ref{fig:13}, for pumping at $2.063$~meV, we obtain a spin-flip only if $\mathcal{K}=-1$, i.e. $K'$ valley is occupied. In case of the valley index $\mathcal{K}=1$, we would not observe any operation on spin for such a driving frequency. This means that we get the spin NOT quantum operation \textit{controlled} by the valley qubit. We denote it simply by $C_\mathcal{K} NOT_s$. For the violet transition ($2.333$~meV) we get the opposite CNOT operation with the spin qubit flipped if $\mathcal{K}=1$, denoted by $\bar{C}_\mathcal{K} NOT_s$. On the other hand, for the green transition we get complementary operation with the valley index being rotated only if the spin is oriented down. In this case acquiring spin-controlled NOT quantum operation on the valley qubit, analogously denoted as $C_{s}NOT_\mathcal{K}$. The purple transition is performed for the opposite, spin-up, thus denoted by $\bar{C}_{s}NOT_\mathcal{K}$. 
Let us note that the CNOT gates are essential in creating the \textit{universal} set of quantum gates. Any (multi-qubit) quantum operation can be approximated by a sequence of gates from a set consisting CNOT gate and some single-qubit operation\cite{shi}, e.g. the $R_{\pi/8}$ gate. 
If we stop our transitions earlier, we can get various rotation gates. In particular, limiting the operation time in Fig.~\ref{fig:14} to $1/4$ of the full valley (spin) flip, i.e. $85$~ps ($230$~ps), we realize the $R_{\pi/8}$ rotation acting on the valley (or spin) qubit.

Beside the both CNOT operations with spin or valley serving as the control qubit, while the other one being the target qubit, we can create previously mentioned SWAPs. By taking the red transition we get spin and valley states swapped, i.e. $|\!-\!1,\uparrow\rangle \leftrightarrow |1,\downarrow\rangle$. The complementary operation, induced by the yellow transition, interchanges the two remaining states: $|\!-\!1,\downarrow\rangle \leftrightarrow |1,\uparrow\rangle$. We denote it by $cSW\!AP$.
All the mentioned two-qubit operations are represented in the spin-valley two-qubit subspace of $|\mathcal{K}, s\rangle$ states by $4\times4$ unitary matrices. Their explicit form can be found in the appendix.

\subsection{Single-qubit gates}

Two-qubit gates are easy to implement here, because the both qubits are specified on two degrees of freedom of the same particle, thus defined in the same localization. Therefore, coupling between them emerges naturally and two-qubit operations require a single transition between one of the four electron basis states. On the other hand, to obtain a single-qubit gate, acting on a given qubit within such a subspace must be done independently from the other qubit state. It turns out that joining two opposite CNOTs makes the operation on the target qubit independent from the control one. If we perform simultaneously both valley-controlled spin NOTs, i.e. $C_\mathcal{K} NOT_s$ and $\bar{C}_\mathcal{K} NOT_s$ we arrive at single spin-NOT quantum gate, denoted as $NOT_s$, independent from the valley degree. Similarly, for simultaneous $C_{s}NOT_\mathcal{K}$ and $\bar{C}_{s}NOT_\mathcal{K}$ we get valley-NOT, $NOT_\mathcal{K}$ quantum operation.

 \begin{figure}[b]
	\center
	\includegraphics[width=8cm]{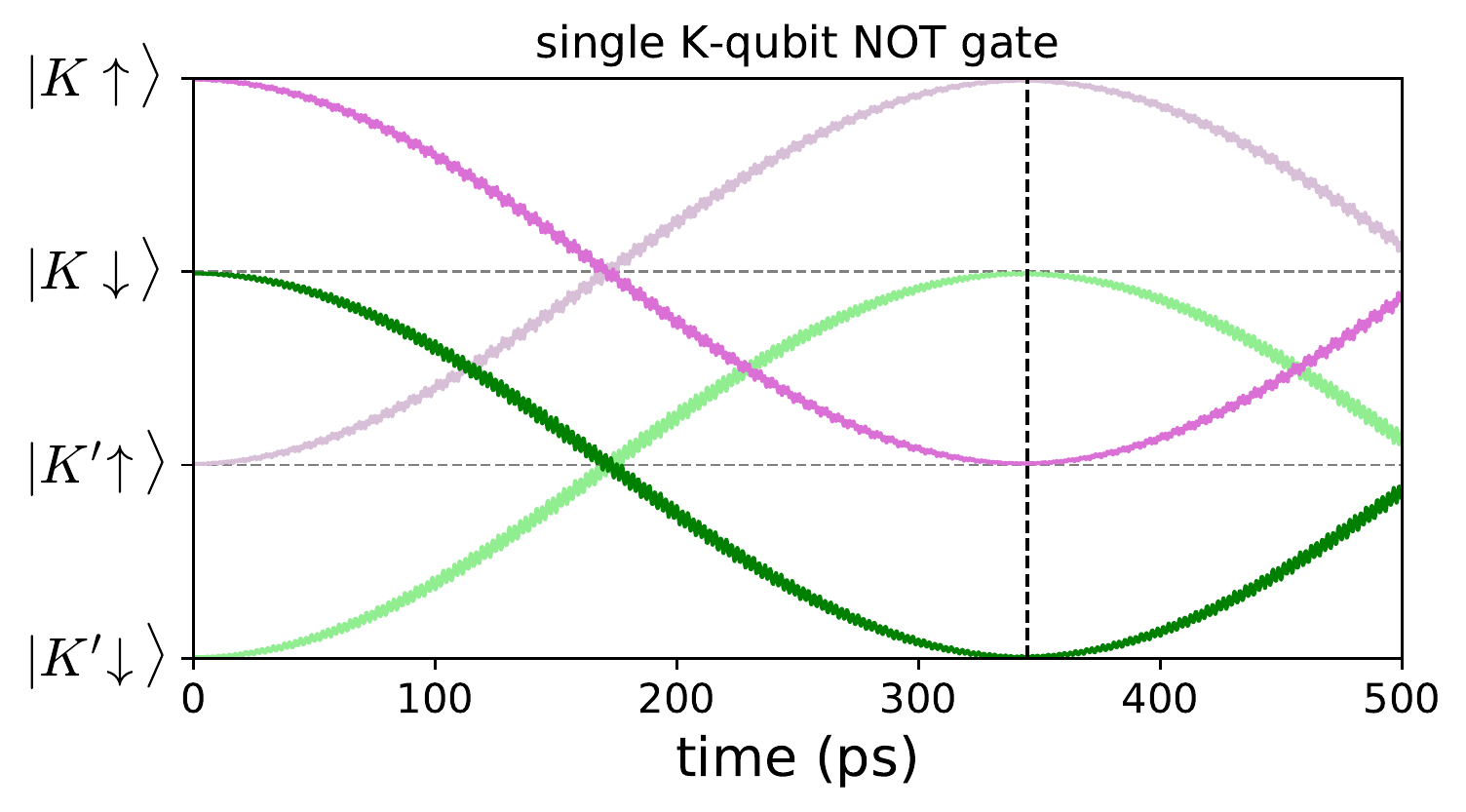}
	\includegraphics[width=8cm]{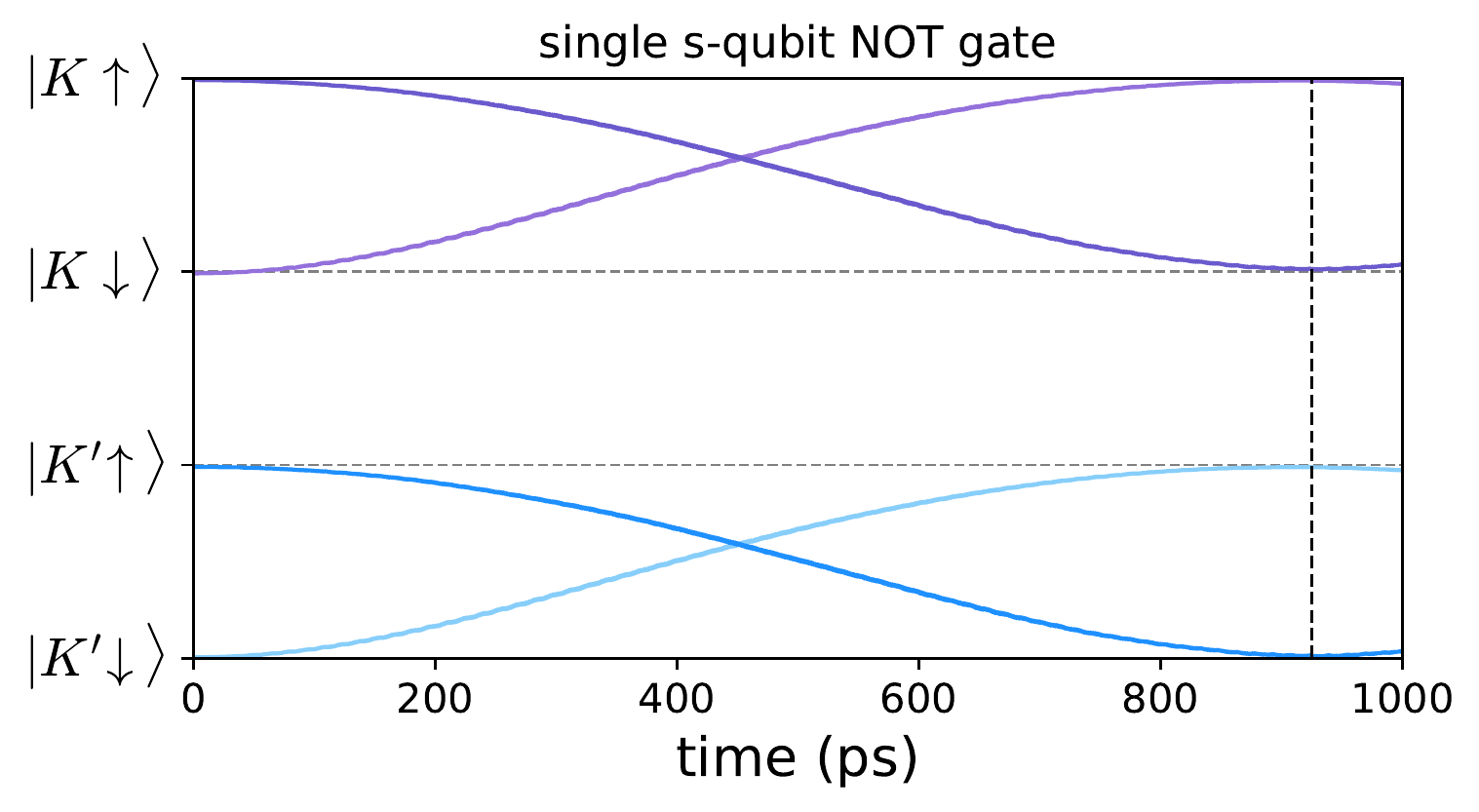}
	\caption{\label{fig:14} Single qubit NOT gate acting on the valley (top) and the spin (bottom) qubit.  Induced simultaneously, both opposite spin controlled valley-flip transitions, marked in Fig.~\ref{fig:13} with green and purple colors, make a single-qubit valley-NOT gate. In a similar way, both valley controlled spin-flips (blue and violet transitions) create the spin-NOT gate.}
\end{figure}
Indeed, to obtain correct operations on spin or valley separately, we need to pump two transitions at the same time. Luckily, it turns out that such twofold transitions are possible, and to do that we need to simultaneously induce oscillations in perpendicular directions, i.e.  $V_{1}(t)=V_{dc}+V_{ac}^x\sin(\omega_x t)$ and $V_{2}(t)=V_{dc}+V_{ac}^y\sin(\omega_y t)$, by feeding both $G_1$ and $G_2$ gates. In Fig.~\ref{fig:14} there are presented twofold transitions which are composed of two intervalley transitions for both spin orientations making up the valley-NOT operation (top: $\mathcal{K}$-qubit NOT gate), and two spin transitions for both valley occupations ($K$ and $K'$) forming $s$-qubit NOT gate (bottom).

The both pumping frequencies $\omega_x$ and $\omega_y$ (blue and violet transitions in Fig.~\ref{fig:14}(bottom)) are very slightly different from these for single separate transitions, e.g. for spin-NOT single-qubit gate they change from $(2.06,2.33)$ to $(2.07,2.35)$~meV for $\hbar(\omega_x,\omega_y)$ respectively.
Whereas the voltage oscillation amplitudes pair $(V_{ac}^x,V_{ac}^y)$ should be selected in a way that the both transitions in the pair lasts the same time. The ratio between them should be properly tuned, e.g. for valley-NOT single-qubit gate (green and purple transitions in Fig.~\ref{fig:14}(top)) $V_{ac}^x/V_{ac}^y=100\mathrm{mV}/62\mathrm{mV}$ for $\hbar(\omega_x,\omega_y)=(1.82,2.58)$~meV respectively.

We see an intriguing feature that local---defined on single electron---two-qubit gates are easier to implement than single-qubit. However, the same single-qubit operations can be obtained a bit easier, simply by performing appropriate CNOTs one by one. Unfortunately, in this simpler approach the operation time is twice as long as for simultaneous twofold transitions. Applying series of operations $C_\mathcal{K} NOT_s\cdot\bar{C}_\mathcal{K} NOT_s= \bar{C}_\mathcal{K} NOT_s\cdot C_\mathcal{K} NOT_s=\mathbf{1}_2\otimes NOT_s$ gives spin-NOT. Similarly,  $C_{s}NOT_\mathcal{K}\cdot\bar{C}_{s}NOT_\mathcal{K}=\bar{C}_{s}NOT_\mathcal{K}\cdot C_{s}NOT_\mathcal{K}=NOT_\mathcal{K}\otimes\mathbf{1}_2$ results in valley-NOT. Both relations can be easily verified by multiplying matrices (included in the appendix) representing the particular operations. 

\subsection{Gate fidelities and qubits readout}

In the course of transitions from Figs. \ref{fig:14} or \ref{fig:12} we can notice a minor oscillation structure of frequency $\hbar\omega$ related to a single cycle of pumping induced by the voltages oscillations. This can be viewed as a reference frame rotation with  $\omega$ frequency in the standard RWA approximation\cite{rwa}. However, it should be emphasized that our numerical calculations are strict. These small oscillations affect the qubit operations fidelity. Fortunately, their amplitude decreases as we get closer to the basis states (i.e. poles on the Bloch sphere). Moreover, we can reduce them arbitrarily by decreasing the amplitude $V_{ac}$ of the voltage oscillations. This is at the expense of increasing the number of pumping cycles, and thus increases the operation time. 
\begin{figure}[b]
	\center
	\includegraphics[width=5.8cm]{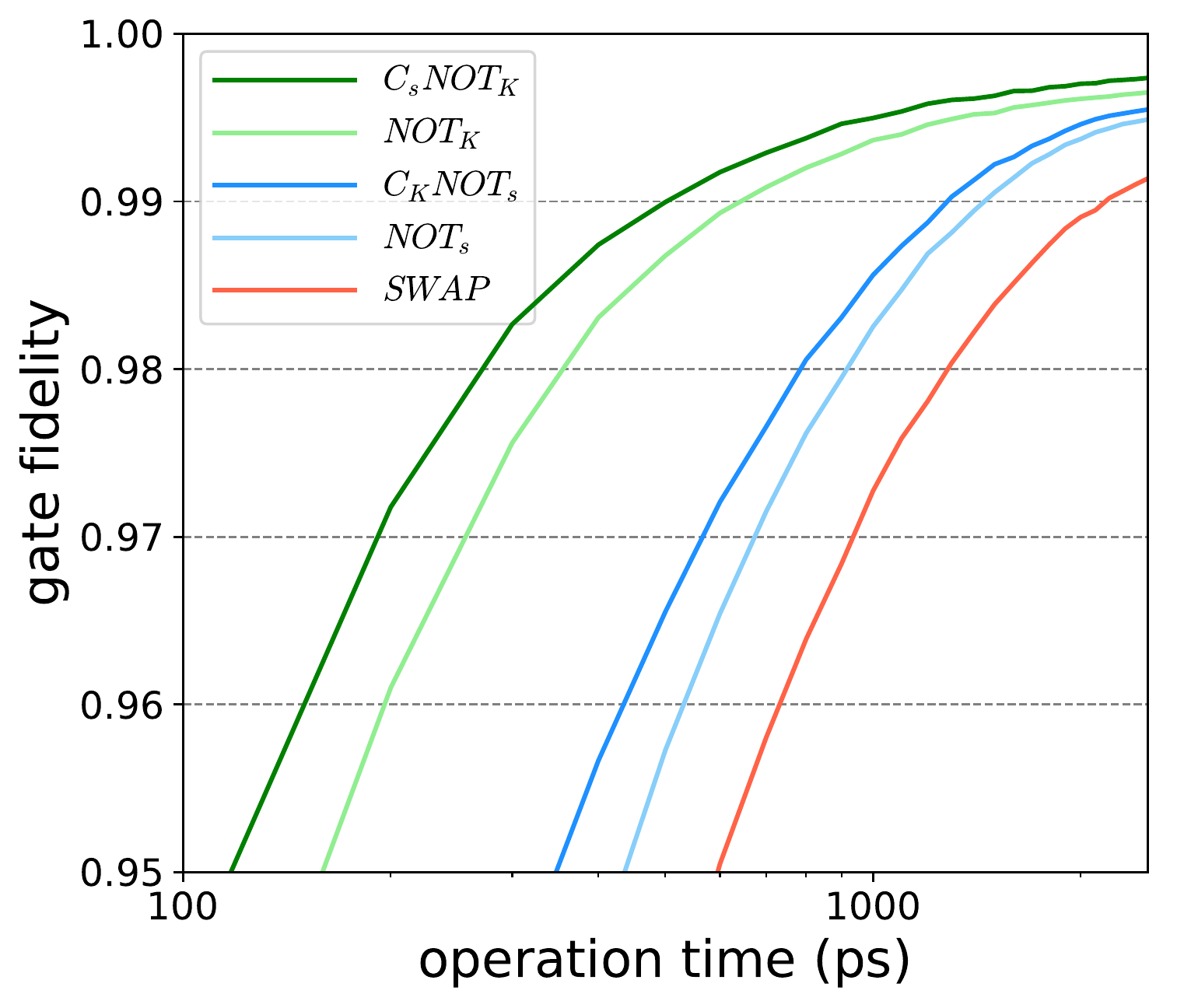}
	\caption{\label{fig:15} Single- (NOT) and two-qubit (CNOT and SWAP) gate fidelities as a function of the operation time. Rising the gates fidelity is done at the expense of increasing the duration of a given operation.}
\end{figure}

We have performed simulations of the operations for decreasing voltage amplitudes, with simultaneously increasing the gates time. In Fig.~\ref{fig:15} there is presented fidelity of various operations as a function of their gate time. One should find an appropriate trade-off between high gate fidelity and low operation time. For example, to obtain an error of the order of 1\% (99\% fidelity) we find valley-related gates ($C_{s}NOT_\mathcal{K}$ and $NOT_\mathcal{K}$) duration of about half a nanosecond, and spin-gates about $1$~ns. The lowest fidelity is for SWAP operations with a $2$~ns optimum for 99\%-fidelity. The duration of operations should be much shorter than the coherence time. 
Estimated\cite{wu} coherence times of electron valley and spin degrees of freedom in MoS$_2$ monolayers, related to the hyperfine interaction decoherence from Mo nuclear spins, are of $100$~ns. This is about two orders of magnitude longer than our operation times. 
The coherence timescale may be slightly overestimated\cite{dec}.
However, it scales with the system size as $\sqrt{N_I}$, with $N_I$ the number of nuclear spins covered by the wavefunction density. Thus, coherence time scales up linearly with the system length, and can be extended by increasing the confinement width.

Each full qubit implementation has to comprise initialization and readout. To do this we can utilize the valley- and spin-Pauli blockade, so far observed in carbon nanotubes\cite{read1,read2}. The blockade utilizes selection rules, which block electron transport between an adjacent dot with the same valley and spin state. Let us add to the setup a nearby auxiliary dot with an electron in the ground state, i.e. $|\mathcal{K}_0, s_0\rangle=|\!-\!1,\uparrow\rangle$. Assuming that valley and spin are conserved during tunneling, the electron carrying our qubits cannot tunnel to the nearby dot if the electron confined there occupies the same spin and valley state\cite{rohling,osika}. The electron is blocked in the same state as its neighbor: $|\!-\!1,\uparrow\rangle(1,1)|\!-\!1,\uparrow\rangle$ and both qubits are initialized. However, when we perform operation on valley or spin qubit, the blockade is lifted and the electron can freely enter the nearby dot: $|\!-\!1,\uparrow\rangle(1,1)|\!-\!1,\uparrow\rangle\rightarrow (0,2)|\mathcal{K}_1, s_1\rangle|\!-\!1,\uparrow\rangle$ with $\mathcal{K}_1=1$ or $s_1=\,\downarrow$. 
In this way, by extending the system with adjacent reference electron, we can perform both---spin or valley qubits readout.

\section{Summary}
The emerging branch of electronics utilizing the valley degree of freedom, called valleytronics in analogy to spintronics, introduces new intriguing methods for defining qubits. Nanodevices with gated monolayer QDs currently become more reliably fabricated. To advantage this, we investigate the possibility of realization of the spin-valley two-qubit system defined on a single electron, that is confined in a QD, controlled by voltages applied to the device local gates. The proposed nanodevice is modeled after structures that were experimentally realized.

A realistically calculated QD confinement potential and electric field via the Poisson equation together with the exact form of the Rashba coupling within the tight-binding monolayer model, leads to reliable modeling of both the intervalley coupling and the Rashba SOI. We solve the time-dependent Schr\"{o}dinger equation with such variable confinement, and track the transitions by calculating actual values of the spin and valley index. We also analyze the edge influence on the intervalley coupling with the increasing flake size and confinement depth,
concluding that proposed nanodevice will work even if we enlarge the flake and move its edges away.

As a result of the performed simulations, we show feasibility of electrically controlling both the electron spin and valley degrees of freedom, simultaneously. By applying an appropriate magnetic field we get such spin and valley Zeeman splittings, that all of the six possible transitions within the spin-valley subspace can be separately addressed. These transitions are interpreted as a variety of two-qubit gates (i.e. CNOTs and SWAPs), and properly combined, they give single-qubit NOT gates. 

Encoding two qubits locally on two degrees of freedom of a single electron reverse difficulty in such a way that two-qubit gates are easer to implement than a single qubit. The latter, however, can be achieved in one go or as two consecutive transitions. By examining the exact course of transitions, we can also estimate fidelity of the implemented gates.

Finally, we remark that to implement fully scalable system, we also need to control interaction between valley (and spin) indexes of nearby electrons in the register. This requires adding additional electrons to the system and research interactions among them.

\begin{acknowledgments}
Author would like to thank Grzegorz Skowron and Pawe\l{} Potasz for invaluable discussions.
This work has  been supported by National Science Centre, under Grant No. 2016/20/S/ST3/00141. 
This research was supported in part by PL-Grid Infrastructure.
\end{acknowledgments}

\appendix
\vspace{1ex}
\section{Rashba coupling parameters}

To calculate the Rashba coupling parameter matrix we utilize the 11-band model from [\onlinecite{ridolfi}] with five $d$-orbitals in the Mo atom
$d_{z^2}$, $d_{xy}$, $d_{x^2-y^2}$, $d_{xz}$, $d_{yz}$, and six $p$-orbitals for the S atoms, three for top ($t$) and three for bottom ($b$) layers: $p^{t,b}_x$, $p^{t,b}_y$, $p^{t,b}_z$.
We add the atomic spin-orbit interaction of the form taken from [\onlinecite{rol,kos}] with intrinsic parameters $\lambda_\mathrm{Mo} = 0.086$ and $\lambda_\mathrm{S}= 0.052$~eV.\cite{kos} After calculating appropriate Slater-Koster elements we obtain the full Hamiltonian $22\times22$ (together witch spin) [see Appendix B in \onlinecite{ridolfi}], with additional onsite potentials $V_{t,b}$ for the top/bottom layers, serving as parameters. 
Now, it can be expressed in an infinite layer-form as a function of the momentum $H(k_x,k_y)$, simply by substituting each hopping $t_{ij}\rightarrow t_{ij}\exp\left(\imath\mathbf{k}\cdot\mathbf{R}_{ij}\right)$. 

We calculate numeric value of $H(k_x,k_y)$ for different $V_{t,b}=\mp E_z d/2$ around the CB minimum, i.e. for $k = K = (\frac{4\pi}{3a},0)$ (or $k = K' = (\frac{2\pi}{3a},\frac{2\pi}{\sqrt{3}a})$) and energy level $\mathcal{E}=2.22$~eV. 
By utilizing the L\"owdin partitioning technique \cite{low1,low2,low3} we downfold it to a reduced Hamiltonian $6\times6$ within our 3-band model\cite{xiao,pavlo}, used in the simulations.
The Schro\"odinger equation for the full $22\times22$ block Hamiltonian is 
\begin{equation}\label{lowdin1}
\begin{pmatrix}
H_{6\times 6} & H_{6\times 16}\\
H^\dag_{6\times 16}& H_{16\times 16}
\end{pmatrix}
\begin{pmatrix}
\psi_{6} \\
\psi_{16}
\end{pmatrix}=
\mathcal{E}\begin{pmatrix}
\psi_{6} \\
\psi_{16}
\end{pmatrix}.
\end{equation}
We perform downfolding by eliminating $\psi_{16}$ and arrive at representation $H'_{6\times 6}\, \psi_6 = \mathcal{E} \,\psi_6 $ where
\begin{equation}\label{lowdin2}
H'_{6\times 6} = H_{6\times 6} +  H_{6\times 16}\left(\mathcal{E}-H_{16\times 16}\right)^{-1}\!H^\dag_{6\times 16},
\end{equation}
which is equivalent to (\ref{lowdin1}).
Afterwards calculating (\ref{lowdin2}), we finally obtain a numeric value of $H'_{6\times 6}$ as a function of $E_z$.
The vector $\psi_{6}$ is represented in the basis of orbitals $(d_{z^2}, d_{xy}, d_{x^2-y^2})\otimes(\uparrow,\downarrow)$, while $\psi_{16}=(d_{xz},d_{yz}, p^{t}_x, p^{t}_y, p^{t}_z, p^{b}_x, p^{b}_y, p^{b}_z) \otimes(\uparrow,\downarrow)$.

After the procedure of downfolding to the 3-band model we obtain a $3\times3$ matrix representing the Rashba Hamiltonian, indexed by the orbital numbers $\alpha$ and $\beta$, made up of $2\times2$ spin blocks. Each of this block has the form $a_0\mathbf{1}_2+a\sigma_x+b\sigma_y$, which we write down as
\begin{equation}
a_0\mathbf{1}_2+
|e|E_z\gamma_{\alpha,\beta}
\begin{pmatrix}
0 & e^{-i\eta}\\
e^{i\eta} & 0
\end{pmatrix}.
\end{equation}
As expected, resulting specific numeric value $e|E_z\gamma_{\alpha\beta}$ of each block $(\alpha,\beta)$ is proportional to the external electric field $E_z$. Moreover, it is multiplied by the matrix with phase factors $e^{\pm\imath\eta}$ of the form equivalent to $e^y_{ij} \sigma_x - e^x_{ij} \sigma_y$ in the $H_R$ Hamiltonian from (Eg.~\ref{rash}). The phase depends on the hopping direction, however in our calculations it is undetermined and resulting $\eta$ was disordered. 

In this way we obtain the parameter matrix $\gamma^{\alpha\beta}_R=|e|E_z\gamma_{\alpha\beta}$, where 
\begin{equation}
\gamma_R=|e|E_z\gamma,\quad
\gamma=
\begin{pmatrix}
	0.09 & 5.62 & 17.4\\
	5.62 & 3.59 & 1.26\\
	17.4 & 1.26 & 4.97
\end{pmatrix}
\times 10^{-3}\;\mathrm{nm}.
\end{equation}
We get the same results for the two remaining $K$ points. However, values obtained around $K'$ point are slightly different
\begin{equation}
\gamma=
\begin{pmatrix}
0.08& 4.58 & 16.9\\
4.58& 4.65 & 0.71\\
16.9 & 0.71 & 3.10
\end{pmatrix}
\times 10^{-3}\;\mathrm{nm},
\end{equation}
Finally, we take to the model an approximate---average value of the $\gamma$ as written down in (\ref{gamm}), remembering, however, of slightly different values between valleys. 
The obtained value corresponds to the Rashba coupling amplitude from [\onlinecite{prx}], where $\lambda_R=3.3\times10^{-4}\,E_z$~(eV nm), for $E_z$ expressed in the $\mathrm{V}/\mathrm{nm}$ units. E.g. for $1$~V/nm and $k$ of order $\frac{4\pi}{3a}$ we get $\lambda_R k\sim 4.3$~meV ($a=0.319$~nm). While in our calculations, for a similar electric field, the $\gamma$ is reaching comparable values of a few meVs.

\section{energy spectrum in magnetic field}
\begin{figure}[h]
	\center
	\includegraphics[width=8cm]{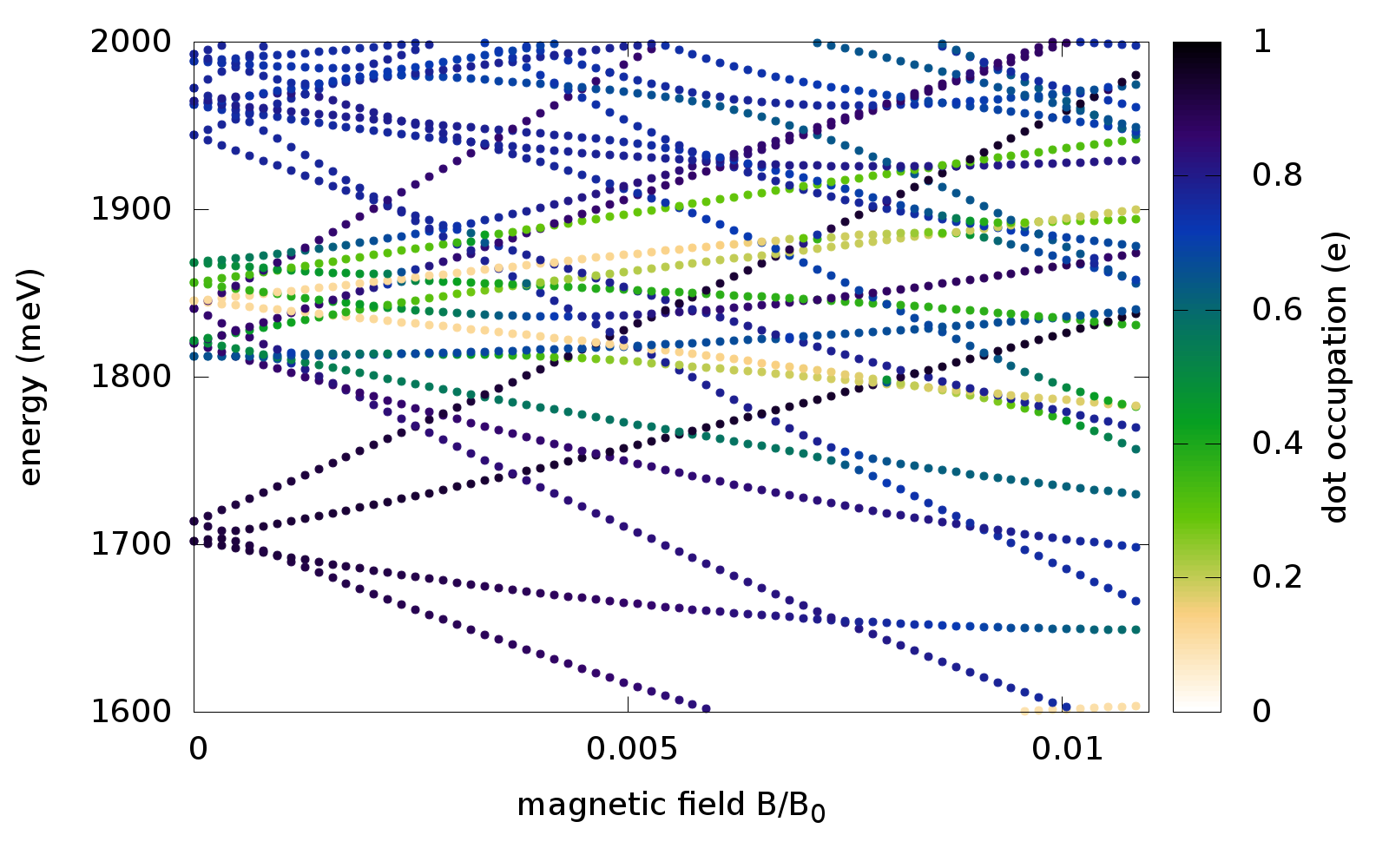}
	\caption{\label{fig:16} The energy spectrum with applied magnetic field ($B_0 = 9.4\times10^4$~T) for an electron confined in the QD. The presented levels comes from the CB minimum.}
\end{figure}
Applying an external magnetic field introduces a splitting of levels with opposite spin (and also valley). If we now gradually increase the magnetic field we obtain typical energy levels structure in  a quantum dot with a magnetic field, presented in Fig~\ref{fig:16}. Similar structures are shown in [\onlinecite{pearce}] (or [\onlinecite{prx}]), calculated within the k.p model for $50$-nm-size (or $40$~nm) MoS$_2$ QD. However here we have more than 10 times smaller dot, thus to obtain similar orbital effects relatively to the magnetic scale $\sqrt{\frac{\hbar}{eB}}$ (size of the Landau ground state), equal $\sim25.66$~nm for $1$~T, we need to increase $B$ field more than 100 times. To keep the results in Fig.~\ref{fig:16} comparable, we omit the Zeeman term here. Presented results are calculated for $N=15$ and $V_{dc}=-1500$~mV, forming the QD confinement.

\begin{figure}[t]
	\center
	\includegraphics[width=8cm]{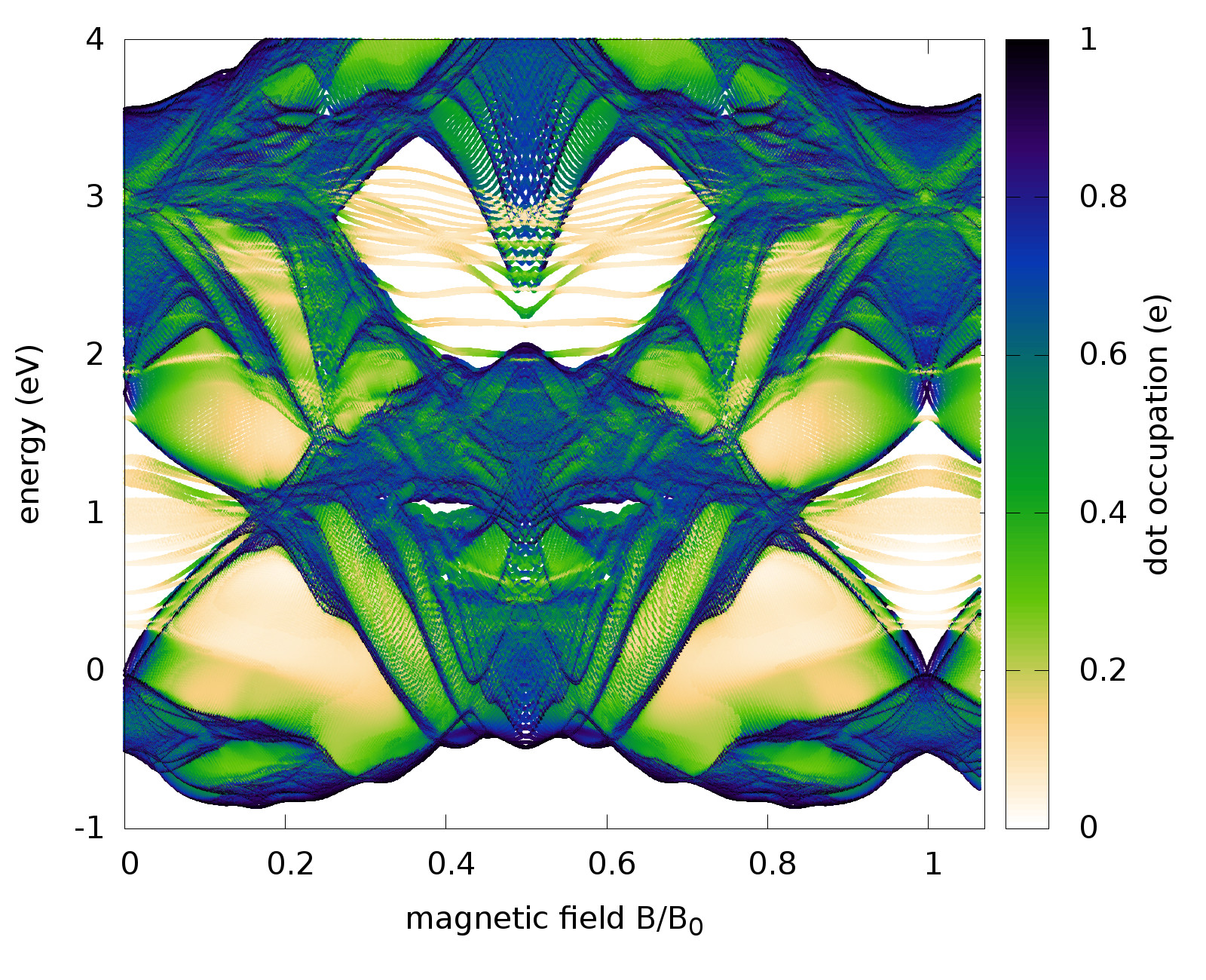}
	\caption{\label{fig:17} The full energy spectrum for electron confined in the QD with applied artificiality strong magnetic field. The energy levels form sophisticated structure called Hofstadter's butterfly.}
\end{figure}

If we further increase the magnetic field, the energy levels will start to attract to each other and form characteristic Landau levels. Calculations for the whole range of artificially high magnetic fields $0<B<B_0$, $B_0 = 9.4\times10^4$~T, shows that Landau levels posses complicated self-similar structure, called the Hofstadter butterfly\cite{hofst,hofst1,hofst2,hofst3}. It is presented in Fig.~\ref{fig:17}. Complex regularities are also manifested in colors that represent the QD occupation. Same here we skip the Zeeman energy term.

\section{two-qubit gate matrices}
Here we present the explicit forms of matrices representing the all two-qubit operations obtained in the simulations. They act in the 4-dimensional Hilbert space of the two-qubit spin-valley states $|\mathcal{K},s\rangle=|\mathcal{K}\rangle\otimes|s\rangle$. The first pair constitutes CNOT operations where the valley is the control qubit and the spin is the target one:
\begin{equation*}
C_\mathcal{K} NOT_s=
\begin{pmatrix}
1& 0 & 0 & 0\\
0& 1 & 0 & 0\\
0& 0 & 0 & 1\\
0& 0 & 1 & 0
\end{pmatrix},\quad
\bar{C}_\mathcal{K} NOT_s=
\begin{pmatrix}
0& 1 & 0 & 0\\
1& 0 & 0 & 0\\
0& 0 & 1 & 0\\
0& 0 & 0 & 1
\end{pmatrix}.
\end{equation*}
While in the second pair, the spin controlled valley-NOT operations are:
\begin{equation*}
C_s NOT_\mathcal{K}=
\begin{pmatrix}
1& 0 & 0 & 0\\
0& 0 & 0 & 1\\
0& 0 & 1 & 0\\
0& 1 & 0 & 0
\end{pmatrix},\quad
\bar{C}_s NOT_\mathcal{K}=
\begin{pmatrix}
0& 0 & 1 & 0\\
0& 1 & 0 & 0\\
1& 0 & 0 & 0\\
0& 0 & 0 & 1
\end{pmatrix}.
\end{equation*}
Finally, the SWAP and its complementary operations are given by: 
\begin{equation*}
SW\!AP=
\begin{pmatrix}
1& 0 & 0 & 0\\
0& 0 & 1 & 0\\
0& 1 & 0 & 0\\
0& 0 & 0 & 1
\end{pmatrix},\quad
cSW\!AP=
\begin{pmatrix}
0& 0 & 0 & 1\\
0& 1 & 0 & 0\\
0& 0 & 1 & 0\\
1& 0 & 0 & 0
\end{pmatrix}.
\end{equation*}
Note that performing jointly both SWAP operations is equivalent to the NOT operation on the both qubits $\sigma_x\otimes\sigma_x$, i.e. $SW\!AP\cdot cSW\!AP = cSW\!AP\cdot SW\!AP = NOT_\mathcal{K} \otimes NOT_{s}$.

\bibliography{spin-valley-qubit}

\end{document}